\documentclass[twocolumn,secnumarabic,amssymb,nobibnotes,aps,prd]{revtex4-2}

\usepackage{setspace}
\usepackage{amsmath} 
\usepackage{amssymb}
\usepackage{mathtools}
\usepackage{amsfonts}
\usepackage{tikz}
\usetikzlibrary{positioning,quantikz}
\usepackage{bm}
\usepackage{bbm}
\usepackage{braket}
\usepackage{graphicx}
\usepackage[style=base,justification=raggedright,singlelinecheck=false]{caption}
\usepackage{subcaption}
\usepackage{url}
\usepackage{color}

\usepackage{makecell}
\usepackage{multirow}

\usepackage{upgreek}
\usepackage{tabularx}
\usepackage{units}
\usepackage{hhline}
\usepackage{textfit} 

\usepackage{dcolumn}
\usepackage{hyperref}
\usepackage[mathlines]{lineno}

\usepackage{indentfirst}
\usepackage[title]{appendix} 
\usepackage{flushend}

\graphicspath{{figures/}}

\begin{document}


\title{Hybrid Classical-Quantum Autoencoder for Anomaly Detection}

\author{Alona Sakhnenko$^{1,2}$}
\email{alona.sakhnenko@tum.de}

\author{Corey O'Meara$^{1}$}
\email{corey.o'meara@eon.com}
\author{Kumar J. B. Ghosh$^{1}$}

\author{Christian B.~Mendl$^{2}$}
\author{Giorgio Cortiana$^{1}$}
\author{Juan Bernabé-Moreno$^{1}$}
\affiliation{$^1$E.ON Digital Technology GmbH, Tresckowstrasse 5, 30457 Hannover, Germany\\
$^2$Technical University of Munich, Department of Informatics, Boltzmannstra{\ss}e 3, 85748 Garching, Germany}
\date{\today}




\begin{abstract}
We propose a Hybrid classical-quantum Autoencoder (HAE) model, 
which is a synergy of a classical autoencoder (AE) and a parametrized quantum circuit (PQC) that is inserted into its bottleneck. The PQC augments the latent space, on which a standard outlier detection method is applied to  search for anomalous data points within a classical dataset. Using this model and applying it to both standard benchmarking datasets, and a specific use-case dataset which relates to predictive maintenance of gas power plants, we show that the addition of the PQC leads to a performance enhancement in terms of precision, recall, and F1 score. Furthermore, we probe different PQC Ansätze and analyse which PQC features make them effective for this task.

\end{abstract}

\maketitle




\section{\label{sec: Introduction}Introduction}


Quantum machine learning (QML) is an emerging interdisciplinary field, which blends quantum computing and machine learning methods to create more efficient schemes for processing data. In this work, we concentrate on a branch of QML that is often referred to as quantum-assisted machine learning \cite{Schuld2018}. Its main goal is to investigate how one can leverage quantum computing to tackle topics that remain challenging for classical machine learning.

Machine learning has demonstrated astounding results with supervised methods, however, for unsupervised learning there remain many open questions and challenges. An important problem in this field is anomaly detection, which prime concern is to identify data points that are uncharacteristic for a specific dataset. The complexity of this task originates in the fact that anomalies are heterogeneous and remain unknown until they occur, and therefore they cannot be formally described~\cite{deep_anomaly}. A popular model for anomaly detection is an autoencoder, which learns a compact representation of the healthy data \cite{10.1145/3097983.3098052}. If there is a failure in detection of a particular data instance, there is a high probability that it encountered something unexpected. i.e., an anomaly. 

Significant effort within the QML community has been devoted to investigate how one can harvest the power of quantum computers that are available today. These machines are small scale ($50-100$ qubits), with limited qubit connectivity, exhibit high error rates, and are known as Noisy Intermediate-Scale Quantum (NISQ)~\cite{preskill2018quantum} devices. Many approaches for NISQ era computers \cite{circuit_learning, transfer, funcke2021dimensional} concentrate on hybrid classical-quantum methods that advantageously combine the strengths of both classical and quantum devices. Autoencoders are well suited for integration with a quantum circuit due to their bottleneck-like architecture. They compress their input, which enables passing high-dimensional classical data into small NISQ computers.

In this article, we present a Hybrid classical-quantum Autoencoder (HAE) model that performs anomaly detection on classical data and is executable on NISQ devices. The HAE consists of a classical encoder, a parametrized quantum circuit (PQC)~\cite{pqc} acting on its output, and a classical decoder, which then processes the measured output of the PQC. The model is trained to reconstruct its input, and, therefore, it initially learns what is normal for the dataset under study \cite{deep_anomaly}. Once the model is trained, we employ the classical encoding layers and the PQC to process the data, which is then analysed by a standard outlier detector - an Isolation forest~\cite{isolation_forest}. The experiment was initially conducted on high-dimensional multivariate sensor data that originated from a gas turbine (see Section~\ref{subsection:principal_dataset}), and then confirmed on further public datasets (see Section~\ref{subsection:additional_dataset}). The results have shown that the way PQC embeds data into the latent space is beneficial for the Isolation forest, and therefore enables more reliable anomaly detection. The effectiveness of a PQC depends on its architecture. We therefore study the effects of over 30 different PQCs and  investigate which features of PQCs are desirable for this task by performing a detailed analysis connecting different measures of model expressivity to its performance. In addition, we explore the relation between the PQC's ability to embed data into a higher dimensional space and the overall performance, although there is no one to one correlation. In the next section, we describe some related works regarding classical and quantum Autoencoders.

\section{Related work}\label{sec:Related work}
\textit{Pang et al.}~\cite{deep_anomaly} provide an extensive survey
of current classical methods for flagging anomalies using deep learning methodologies. One popular approach makes use of Autoencoders (AEs) \cite{baldi2012autoencoders, makhzani2015adversarial} or its more powerful variant - Variational Autoencoders (VAEs) \cite{kingma2019introduction, doersch2016tutorial, sonderby2016ladder}. This type of models serve as a foundation to this project.

Initial work on quantum autoencoders focused on utilizing their bottleneck architecture for compression of quantum data \cite{Romero_2017}. The authors showed an application of this idea in the context of quantum simulations. However, in the present work, we would like to concentrate on classical data being processed by a quantum algorithm.

\textit{Herr et al.}~\cite{qGAN4anomaly} applied quantum generative models for anomaly detection on classical data. In their pioneering work, the authors utilized a network architecture similar to AnoGAN~\cite{anogan} and introduced a hybrid classical-quantum structure for the generator network. However, in their work, classical data is only passed into the discriminator, therefore a pre-processing layer for classical data in the generator is not needed. Even though this approach did not show any advantage in performance compared to its classical counterpart on the chosen dataset, it laid an important groundwork for further research.

One of the first studies to investigate the correlation between quantum model classification accuracy and expressivity descriptor as proposed \textit{Sim et al.}~\cite{expressibility_haar} was conducted by \textit{Hubregtsen et al.}~\cite{similar_paper}. The experiment was performed on purely quantum models and on artificial classical data, which showed strong correlation between classification accuracy and expressivity measure. In our work, we build up on this study and test these descriptors in a hybrid classical-quantum scenario on a real-world dataset \cite{gircha2021training, pramanik2021possible, srikumar2021clustering}.

\section{Problem Description}\label{section:Problem Description}

In this section, we detail our approach for anomaly detection in the latent space of an autoencoder, introduce the performance metric used, and list datasets that our model was tested on. 

\subsection{Detecting outliers in the latent space}\label{subsection:anomalies_in_latent}

How can we identify anomalies with an autoencoder? The most straightforward way would be to look for unexpectedly high reconstruction loss \cite{asperti2020balancing, sabokrou2016video}. This will indicate that a model encountered something that it has not seen before. Alternatively, we can assume that anomalies are embedded in the latent space in unexpected places. Through empirical trials, we have determined that this method delivers a more reliable result. Therefore, to flag an anomaly, we extract latent space and apply \textit{Isolation forest}~\cite{isolation_forest} algorithm to mark anomalous instances. The core idea behind this method lies in an assumption that outliers are sparse and contrasting to the rest of the data, and hence easier to isolate. 

To gauge the performance of the algorithm, an accuracy metric is required. For this purpose, we choose \textit{precision} (fraction of true anomalies of all discovered instances), \textit{recall} (fraction of true anomalies that were discovered) and a harmonic mean of them known as the \textit{F1 score}~\cite{f_score}. These metrics allow us to evaluate the performance of the model on imbalanced datasets. Given that some datasets in this project contain time-series data, we allow the algorithm to detect an anomaly within a certain time interval of a ground truth data point. Although  \textit{F1 score} has two limitations, namely, not normalized and not symmetric,  but for our purpose these limitations don't impact the classical vs hybrid Autoencoder comparison.


\subsection{Gas turbine dataset}\label{subsection:principal_dataset}

\begin{figure}[htb]
    \centering
    \includegraphics[width=0.5\textwidth]{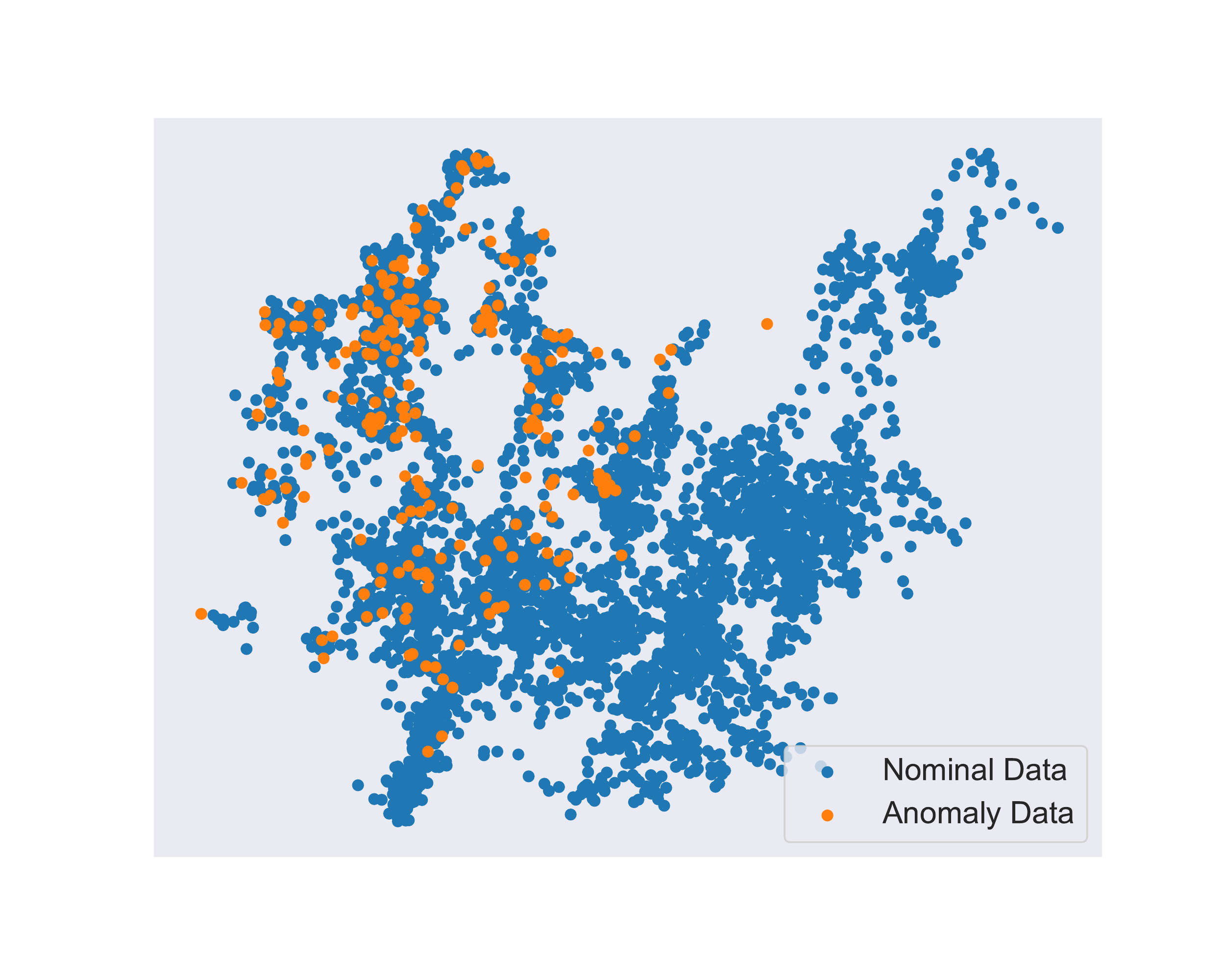}
    \caption{Visualization of the Gas turbine dataset with anomalies (orange) and nominal data (blue). Each data point represents a single time $t$ of all $161$ measured sensor values. For visualization purposes, Principal Component Analysis was performed to reduce the dimension to $4$ (plotted here as 2-dimensional subspace). The anomaly data points correspond to times when measured alarms occurred on the asset. Section~\ref{sec:final_arch_hae_gas_turbine} describes the final anomaly detection results for this difficult benchmark dataset.}
    \label{fig:gas_turbine_benchark}
\end{figure}

We consider a real world scenario for a predictive maintenance system of a power plant. The power plant contains a gas turbine which has $161$ different sensors which record measurements such as: input gas pressure, bearing temperatures, water coolant temperatures, and various other measurements. The $161$ sensors were recorded for a duration of $10$ months with granularity of $10$ minutes. For this analysis, the dataset was reduced to use $640$ data points for training data (equal to approximately $2$ weeks of sensor data) and $2000$ points for testing (equal to approximately $4$ weeks of time-series sensor data). See Appendix \ref{section:data_proprocessing} for data pre-processing details and Fig. \ref{fig:gas_turbine_benchark} for dataset visualization.

What makes the problem difficult is that this real life scenario is an unsupervised classification task where no explicit times have been flagged by industry experts as representing anomalous behaviour of the plant. While the gas turbine was running in operations for a period of $1$ month, another monitoring system was able to capture hundreds of alarm triggers occurring at various times. We use these alarm times as rough labels for anomaly data points. These labels consist of over $200$ different types of alarms and shutdowns. We reduce the amount of alarms by changing their granularity to $10$ minutes. The types of alarms are not taken into account, and all alarms and shutdowns are treated the same. An important point is that not all alarms which were triggered may be correlated with the operational state of the gas turbine. Therefore, it is not always the case that at a certain time, where we are assuming an anomaly occurred, there should exist an appropriate effect in the measured data at that time. Figure~\ref{fig:gas_turbine_benchark} demonstrates that many anomaly data points are indeed straying away from central dense clusters however, some data point do not appear as outliers.

\subsection{Additional public datasets}\label{subsection:additional_dataset}

To demonstrate the robustness of our new approaches, we benchmark similar publicly available classification datasets which have been used in outlier and anomaly detection studies \cite{Rayana_2016}. These classification datasets have been framed as an outlier detection problem for an unsupervised learning setting. Each data point has a pre-defined label of \texttt{0} or \texttt{1} to indicate whether the point is an inlier or an outlier. The three public datasets used in this project are the following:

1. The Musk dataset \cite{aggarwal_musk_2015,jain1994compass} that consists of molecule data with 166 features. The model has to learn to separate non-musk molecules (\texttt{class~0}) from the musk ones~(\texttt{class~1}). 

2. The Arrhythmia dataset \cite{liu2008isolation,guvenir1997supervised} that contains heart measurement data with 274 features. The goal is to differentiate healthy data (\texttt{class~0}) from data that indicates one of the types of cardiac arrhythmia (\texttt{class~1}). We do not differentiate between different classes of arrhythmia and treat them as a single outlier class.

3. The Statlog (Landsat Satellite) database \cite{liu2008isolation} that consists of satellite image data with 36 features. This is a multi-class classification dataset with 7 classes depending on the soil colors in the satellite images. The smallest three classes were merged to form \texttt{class~1}, while the rest represent the healthy data (\texttt{class~0}).

Like our principal dataset, the Musk and Statlog contain medium to high dimensional well clusterable data, whereas Arrhythmia has more uniform data distribution in its high dimensional feature space. 

\section{Hybrid Classical-Quantum Autoencoder (HAE)} \label{section:hybrid_autoencoder}

\begin{figure*}[t]
    \centering
    \includegraphics[width=\textwidth]{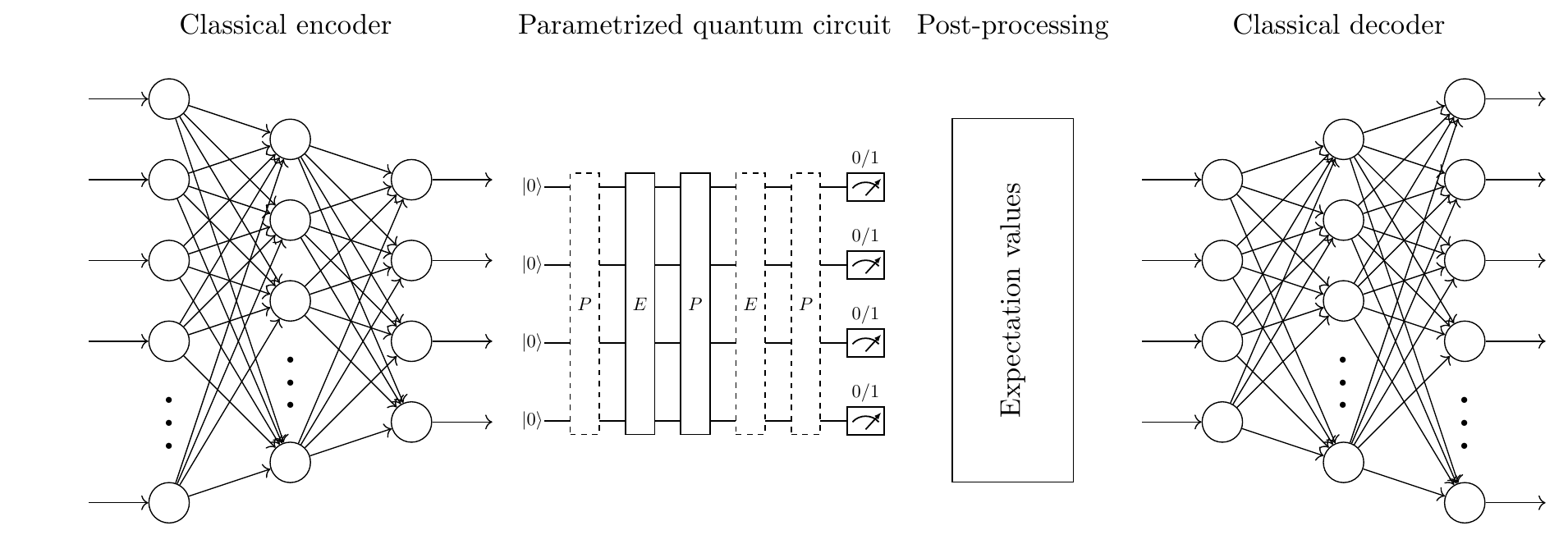}
    \caption{An illustration of the main building blocks of a hybrid classical-quantum autoencoder. First, the data is compressed by the classical encoder to the size of 4 units. This information is then embedded into the 4-qubit quantum circuit by the encoding layer, denoted here as E, and processed by the processing layer P. E and/or P layers can be repeated, based on the chosen circuit architecture. The dotted boxes within a PQC represent optional layers. The final quantum state is then measured with respect to the (standard) $Z$-basis, and a corresponding expectation value for each qubit is calculated. These expectation values span the latent space that is used to detect anomalies. Afterwards, the decoder expands the compressed data representation to its original size.}
    \label{fig:hybrid_autoencoder}
\end{figure*}

In this section, we introduce the HAE model, its training details and analyse the impact of different PQC architectures on the overall performance of the model. In the proposed HAE architecture, we aim to leverage the strengths of both classical and quantum machines. We utilize the memory of the classical machine to load and compress the data, which is then processed by a quantum circuit. A sketch of our proposed architecture can be seen in Figure~\ref{fig:hybrid_autoencoder}. As it can be seen, a PQC consists of embedding and processing layers. We concentrate on embeddings that are realizable with parametrizable gates that are summarized in Figure~\ref{fig:embeddings}. Pre- or post-processing steps are implemented following quantum neural network architectures that consist of parametrizable, non-parametrizable and entangling elements organized in layers. The outputs of a PQC are Pauli-Z expectation values for each qubit.

\subsection{Gas turbine anomaly detection problem}\label{subsection:benchmark}

The Gas turbine dataset is described in Section~\ref{subsection:principal_dataset} and its associated anomaly detection problem is to determine which times an anomaly occurred for an operational gas turbine used in a power plant. To examine the effects of PQCs on model efficiency, we need to first test the performance of classical models without augmentations. The autoencoder (AE) is implemented in \texttt{PyTorch}~\cite{pytorch}. An encoder consists of three \texttt{Linear} layers of dimensions \texttt{input size}, \texttt{56} and \texttt{4} (latent space) with \texttt{tanh} activation functions. The decoder is a mirrored encoder. 

\begin{figure}[htbp]
     \centering
    \begin{subfigure}[b]{0.5\textwidth}
         \centering
         \includegraphics[width=\textwidth]{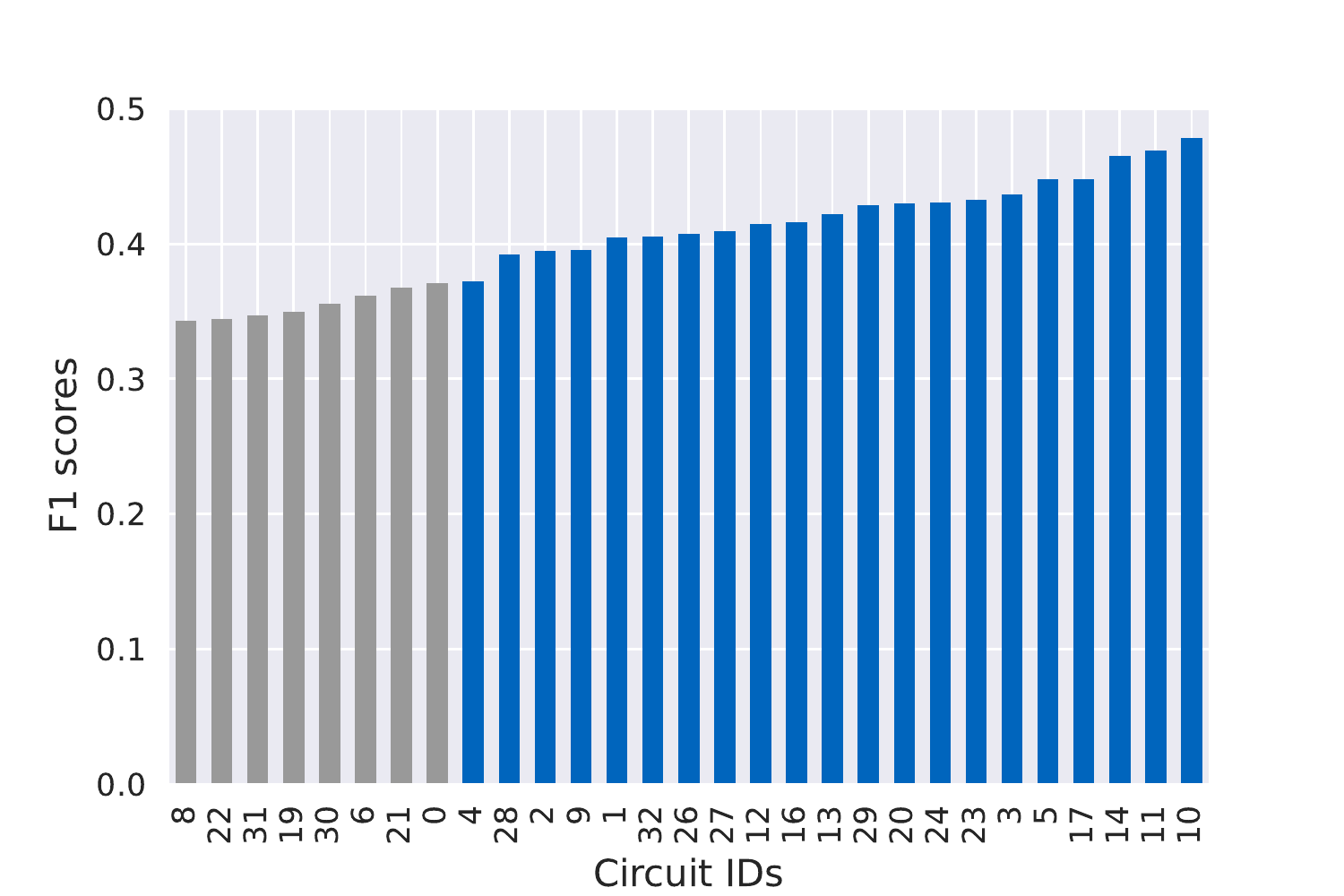}
     \end{subfigure}
     
     \begin{subfigure}[b]{0.5\textwidth}
         \centering
         \includegraphics[width=\textwidth]{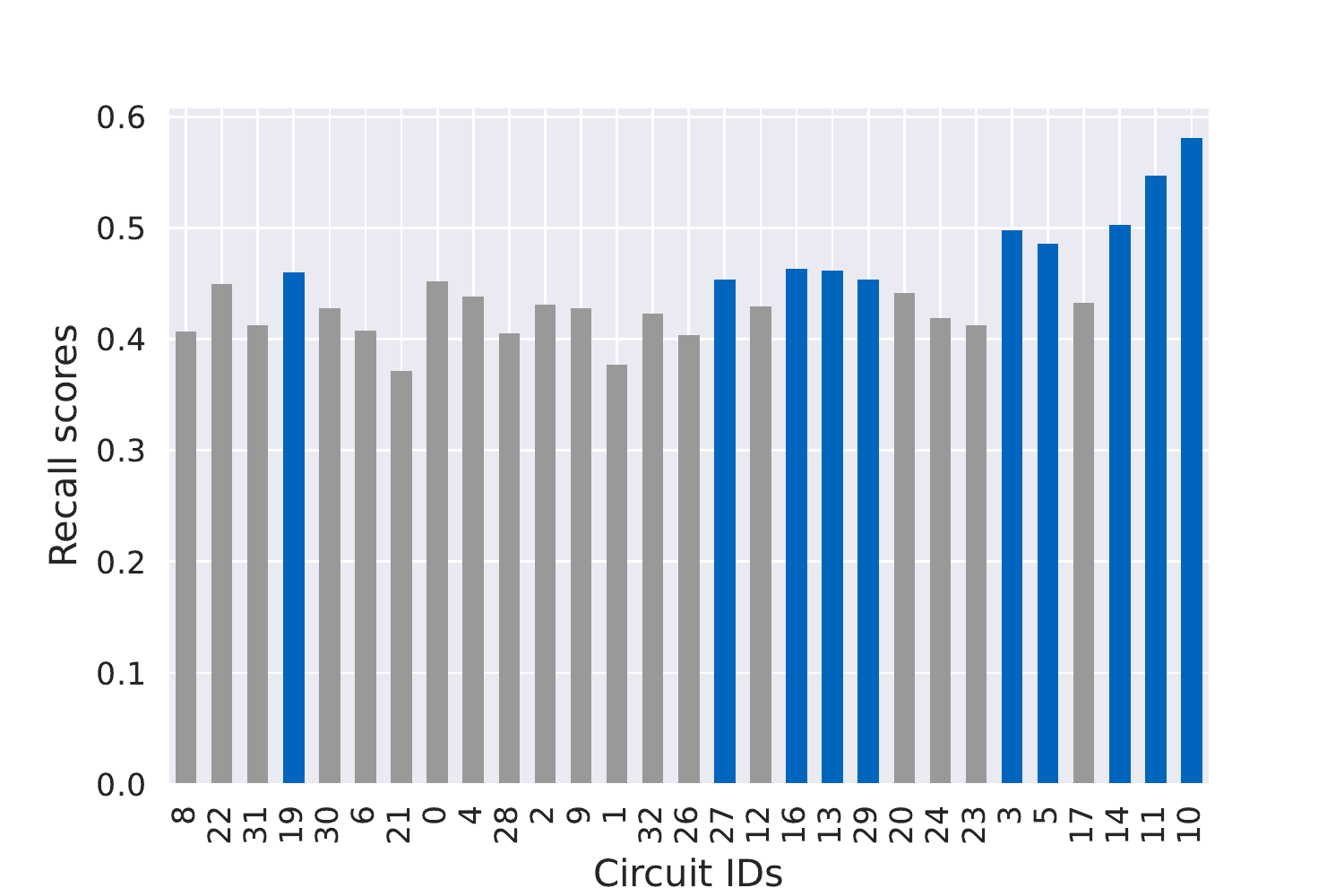}
    \end{subfigure}
     
    \begin{subfigure}[b]{0.5\textwidth}
         \centering
         \includegraphics[width=\textwidth]{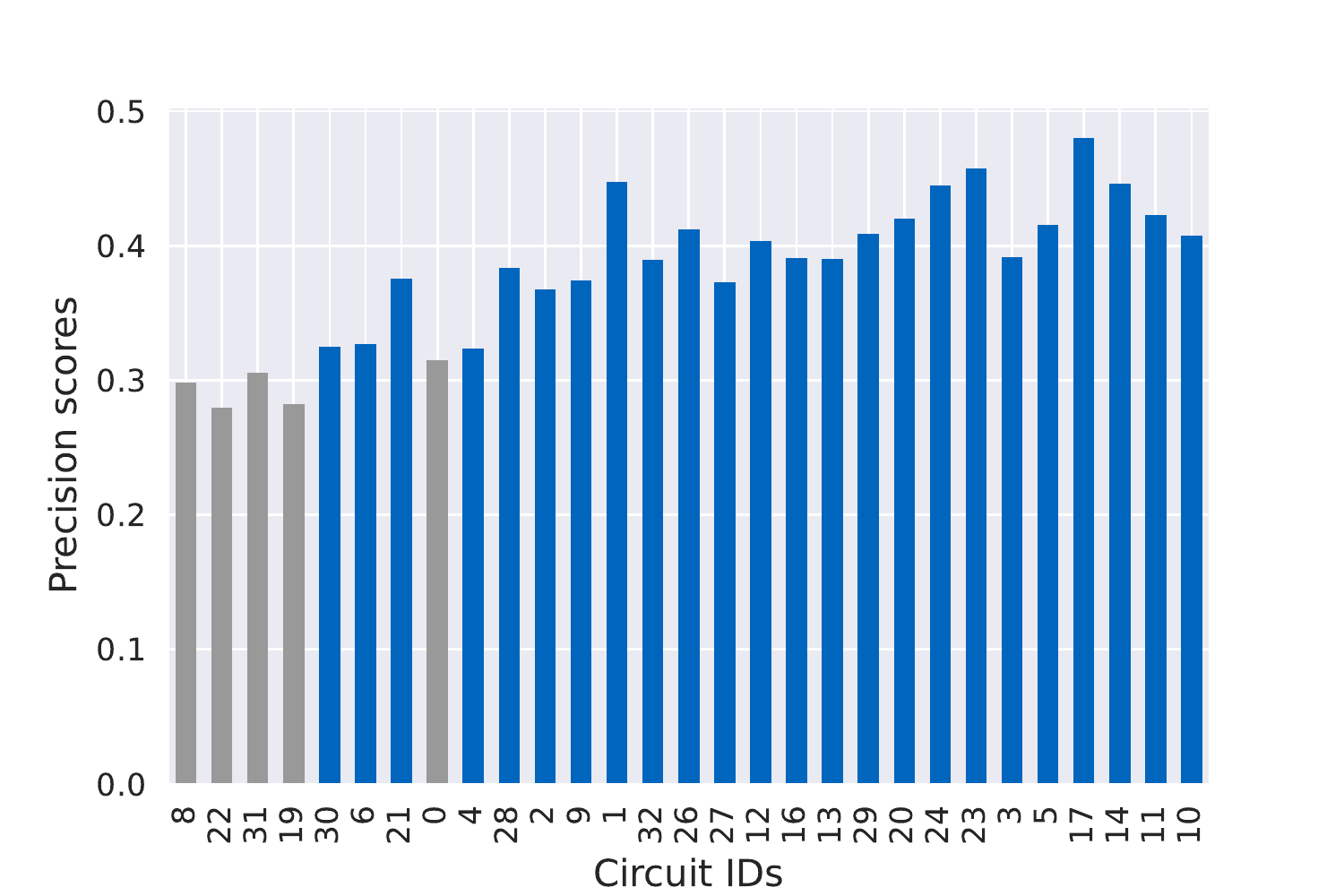}
    \end{subfigure}
     \caption{Performance comparison of different quantum circuit models. The x-axis represents different circuit IDs, descriptions of which can be found in Appendix~\ref{section:circuit_list}. The blue colour indicates scores that are higher than classical AE, which is indicated here as circuit with ID~0.}
     \label{fig:qae_performance}
\end{figure}

The testing method of the model is the following: (1) we train the AE with \textit{mean square root reconstruction error} for \texttt{80} epochs, batch size of \texttt{32}, and learning rate set to \texttt{0.001} on \texttt{640} training data points which corresponds to approximately $1$ week of data; (2) once the AE has been trained, we use the encoder to reduce the dimension of the complete test dataset to \texttt{4}; (3) we then train and test a \texttt{sklearn.ensemble.IsolationForest} model on this new dimension reduced data in order to determine the quality of the dimension reduction performed by the autoencoder. Each model was trained three times, and we recorded the average performance for all three training iterations. The resulting precision, recall and F1 score of this approach are shown in Figure~\ref{fig:qae_performance} as Circuit 0.

We can now move to the implementational details of HAE. The general architecture of HAE (see Figure~\ref{fig:hybrid_autoencoder}) was implemented by augmenting the previously developed \texttt{PyTorch} autoencoder with a quantum circuit that is implemented using \texttt{Qiskit}~\cite{Qiskit}. All PQCs are \texttt{4-qubit} systems that are simulated using the \texttt{IBM statevector simulator}. We preserve the same training approach and hyperparameters described above, except that we use both a classical encoder and a PQC to process data that is then analysed by the Isolation forest. We probe over 30 different circuits design to see which provide a boost to our model's performance. The list of all attempted circuits can be found in the Appendix \ref{section:circuit_list}, they were designed in a similar vein as the ones listed in \cite{expressibility_haar}. The resulting precision, recall and F1 score of all converged circuits are shown in Figure~\ref{fig:qae_performance}. The bars that are coloured blue highlight circuit IDs that show a higher score than the classical model.

In the next section, we investigate which variations of quantum circuits have an impact on the overall anomaly detection measures and infer some interesting relationships between different quantum circuit metrics.

\subsection{PQC descriptors}\label{subsection:analysis_descriptors}

How can we make a meaningful design choice for PQCs? Often they are designed by domain experts for specific tasks. Here, we explore an alternative approach, in which we consider different descriptors that characterize behaviours of PQCs, and we analyse how they affect models' performance. 

A common way to characterize a circuit in the literature is by the amount of trainable parameters $\theta$. However, this metric is too general and does not capture enough about model's expressibility. In this article, we investigate how the quantum circuit descriptors proposed by \textit{Schuld et al.}~\cite{encoding_fourier} and \textit{Sim et al.}~\cite{expressibility_haar} correlate with the behaviour of the quantum model $\mathcal{M}_{\theta}$. These descriptors can be summarized in the following fashion:
\begin{enumerate}
    \item Estimating the divergence between the fidelity density $F$ of generated states from the quantum circuits and the fidelity of Haar random states, defined as
\begin{equation}\label{eq:expressivity_haar}
    \text{Expr}_{\text{Sim}}(\mathcal{M}_{\theta}) = \mathbb{KL}(P_{\mathcal{M}_{\theta}}(F)|| P_{\text{Haar}}(F)),
\end{equation}
where $P_{\text{Haar}}(F) = (d - 1)(1 - F)^{d - 2}$ for a $d$-dimensional Hilbert space. This measure represents how much the model ``explores'' the associated Hilbert space and was proposed by \textit{Sim et al.}~\cite{expressibility_haar}. Henceforth, this measure is referred to as \textit{Sim expressivity}. It calculates the KL divergence between $P_{\text{Haar}}(F)$ and $P_{\mathcal{M}_{\bm{\theta}}}(F)$. Conveniently, $P_{\text{Haar}}(F) = (d - 1)(1 - F)^{d - 2}$ has already a closed-form solution. To compute $P_{\mathcal{M}_{\bm{\theta}}}(F)$ we need to first calculate fidelities of randomized circuits. In \cite{expressibility_haar}, embedding was neglected, however, in this article, embedding constitutes an essential part of the circuit and hence will also be considered regarding the expressivity of the circuit. We sample $\theta \sim U[-\pi, \pi]$ from uniform distribution to encompass values on the entire Bloch sphere and $\mathbf{x} \sim U[-1, 1]$ to cover all values within the input range (due to \texttt{tanh} activation functions). We draw \texttt{1000} samples for each circuit. To compute a discreet KL divergence, we use a histogram and select discretization of \texttt{100} bins.
    
    \item Measuring the entangling capacity. This can be done with the Meyer-Wallach measure \cite{expressibility_haar} that is averaged over a parameter set, or by taking into account the amount of entangling layers. These measures would show how dependent qubits are on each other. In \cite{meyer_wallach_easy} the authors showed that Meyer-Wallach measure is connected to the average purity of qubits and can be formulated alternatively as:
\begin{equation}
\text{Capacity}(\ket{\psi}) = 2 \left(1 - \frac{1}{n} \sum_{k=0}^{n-1}\mathrm{Tr}[\rho_k^2]\right),
\end{equation}
where $\rho_k$ is a density matrix of $\psi$ with a traced out $k$-th qubit. This formulation is easier to implement and hence was employed in this project.
    
    \item Assessing the complexity of functions that a quantum circuit can represent by means of Fourier analysis as proposed by \textit{Schuld et al.}~\cite{encoding_fourier}, who showed that a univariate quantum model can be represented through a partial Fourier series:
    \begin{equation}\label{eq:fourier_series_exp}
    \mathcal{M}_{\theta} = \sum_{\omega \in \Omega} c_{\omega} e^{i \omega x},
    \end{equation}
    where $\Omega$ represents a frequency spectrum and $c_{\omega}$ stands for a coefficient that corresponds to a frequency $\omega$, which can be extended to a multivariate case. Their work showed that the embedding strategy controls the Fourier frequency spectrum $\Omega$, while the structure of the remaining circuit influences the expressivity of the coefficients $c_{\omega}$. The spectrum influences the family of functions that a PQC can represent, and therefore bounds the expressivity of the model. It can be augmented by repeatedly (redundantly) embedding the data into the quantum domain. To gain a better intuition of this method, we run a 1-qubit circuit experiment and experiment with different architectures. The results are presented in Appendix~\ref{appendix_fourier}.
    
    The original paper~\cite{encoding_fourier} provided a mathematical derivation to support their claim, and the authors have mainly concentrated on the Pauli-$X$ embedding strategy for the single-qubit case. To test these ideas on a broader scale of PQCs, we utilize Fourier transformation to extract relevant information for an arbitrary PQC. As expressivity descriptors, we choose the amount of the positive frequency in the spectrum, the amount of available phases  and variance of the amplitudes (see Appendix~\ref{appendix_fourier}). The last two descriptors are meant to capture the expressivity of the Fourier coefficients.

\end{enumerate}

\begin{table*}[t]
    \centering
    \begin{tabular}{l|c|c|c|c}
          \multirow{3}{*}{\textbf{Descriptors}} & \multicolumn{4}{c}{\textbf{Correlation with}}\\
           & \textbf{Precision} & \textbf{Recall} & \textbf{F1 score} & \makecell{\textbf{Reconstruc-} \\\textbf{tion loss}} \\
         \hline
         \multicolumn{5}{c}{All converged PQCs} \\
         \hline
         Amount of parameters & 0.054 & 0.390 & 0.265 &  0.045\\
         Sim expressivity & 0.290 & -0.289 & 0.052 & 0.523\\
         Amount of entangling layers & 0.502 & 0.341 & 0.562 & 0.443\\
         Meyer-Wallach measure & 0.317 & 0.507 & 0.481 & 0.342\\
         Amount of positive frequencies & 0.198 & 0.304 & 0.319 & 0.276\\
         Amount of phases & 0.175 & 0.322 & 0.334 & 0.200\\
         Variance of amplitudes & -0.083 & -0.0405 & -0.105 & 0.0356\\
         \hline
         \multicolumn{5}{c}{PQCs that share Pauli-$X$ embedding} \\
         \hline
         Amount of parameters & 0.270 & 0.439 & 0.427 &  0.259\\
         Sim expressivity & 0.356 & -0.364 & -0.0450 & 0.308\\
         Amount of entangling layers & 0.636 & 0.414 & 0.631 & 0.611\\
         Meyer-Wallach measure & 0.876 & 0.732 & 0.876 & 0.644\\
         Amount of positive frequencies & 0.452 & 0.613 & 0.642 & 0.612\\
         Amount of phases & 0.385 & 0.445 & 0.513 & 0.430\\
         Variance of amplitudes & -0.192 & -0.089 &	-0.186 & 0.134
    \end{tabular}
    \caption{This table summarizes the correlation between different PQC descriptors on precision, recall, F1 score and reconstruction loss. These results were acquired on the principal dataset (see Section~\ref{subsection:principal_dataset}). The first half of the table holds statistics of all 28 converged PQCs (see Appendix~\ref{section:circuit_list}), while the second part only from PQC that share the same embedding, which in this case are Pauli-$X$ embeddings (14 PQCs in total).}
    \label{tab:descriptor_results}
\end{table*}

We examine the correlations between different descriptors and precision, recall, F1 score and reconstruction loss. The collected statistics are summarized in Table~\ref{tab:descriptor_results}. During the analysis of the model training results, we observed an interesting phenomenon as reconstruction loss seems to rise with increase in precision and F1 score (we observe correlation of $\sim0.61$ and  $\sim0.45$ respectively). This is unanticipated as HAE is primarily trained to reconstruct its input and thus learn to represent healthy data as good as possible, while anomaly detection is an auxiliary task. We examine this matter closer below.

Table~\ref{tab:descriptor_results} shows a positive correlation of $\sim0.523
$ between Sim expressivity measure and reconstruction loss on a test dataset. It suggests more expressible circuits are better at reconstructing the input (since for both Sim expressivity and reconstruction loss, lower numbers are desirable). Other descriptors seem to negatively impact the reconstruction loss as we are seeing positive correlation between them (e.g., reconstruction loss is higher with more entangling layers).

This peculiar behaviour seems to indicate that Sim expressivity is the only descriptor that assist HAE during training, which aligns with results in \cite{similar_paper}, while others hinder it. The growth in reconstruction loss with increasing circuit complexity is, however, not surprising and can be ascribed to the fact that training parameters were held static throughout the experiment. This would mean that we are trying to traverse progressively more complex optimization terrain without adapting our strategy. All the parameters remain static throughout the experiment as they are defined in Section~\ref{subsection:benchmark}. However, we observe an increase in F1 score despite declining training success. A plausible explanation for this phenomenon is that quantum circuits still manage to learn enough to position the outliers isolated from the healthy data in the latent space, which is a favourable development for the Isolation forest. This behaviour can be explained by considering this model from a kernel methods perspective, as suggested in \cite{schuld2021supervised}, however, the details of this connection between our model, Isolation forest and kernel methods require further investigation in future work. 

One illustrative example that grounds the more general claim from above is performance comparison of Circuits with IDs 3 (Figure~\ref{circuit_3}), 10 (Figure~\ref{circuit_10}) and 13 (Figure~\ref{circuit_13}). The best performing circuits (Circuit 10) is Circuit 3 twice repeated. Figure~\ref{fig:qae_performance} shows that performance in terms of recall and F1 score grows from Circuit 3 to Circuit 10 as expected. However, if Circuit 3 is repeated thrice, as in Circuit 13, the performance drops significantly. This example highlights that a more complex circuit failed to provide a further boost in performance, possibly due to reasons stated above.

Interestingly, Sim expressivity seems to assist recall, but hinder precision and, like other descriptors, apart from entangling strength, has a low impact on precision, recall and F1 score if all PQCs are considered together. However, if we focus on PQCs with the same embedding strategy, which in this case was angle embedding, the impact of all descriptors grows. Peculiarly, the descriptors that are associated with Fourier analysis (particularly, the amount of positive frequencies and phases) have a big impact on recall and F1 score in this scenario. However, their influence diminishes if all PQCs with different embedding strategies are considered. Variance in amplitudes as a descriptor does not seem to capture the relevant behaviour well.

\subsection{The effects of classical expansion of the latent space} \label{section:quantum layer is replaced by classical layers}

\begin{figure*}[htb]
    \centering
    \includegraphics[width=0.75\textwidth]{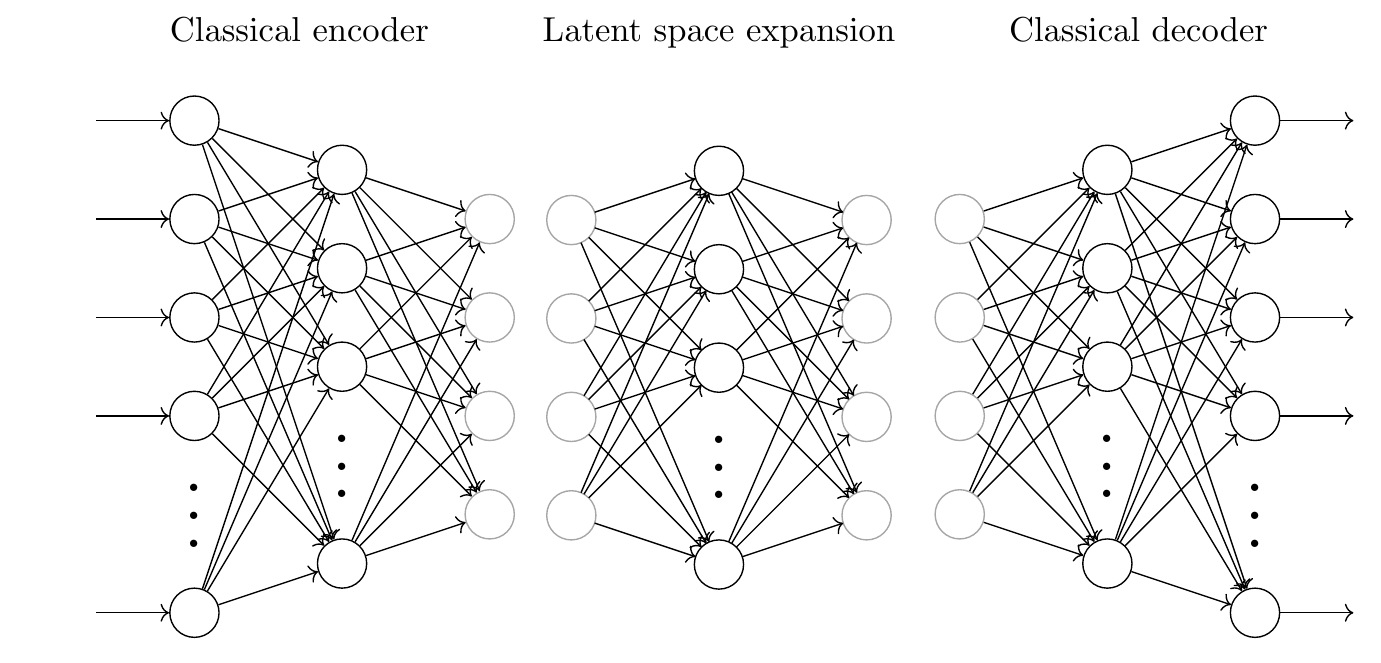}
    \caption{An architectural sketch of the autoencoder with a classically expanded latent space. In this model, PQC and post-processing part of the model (see Figure~\ref{fig:hybrid_autoencoder}) have been replaced by an additional $16$-dimensional layer that mimics the dimensionality of the Hilbert space of the PQC. The output from the latent space expansion insertion is used to detect anomalies. The greyed out nodes depict the same nodes.}
    \label{fig:newarchitecture}
\end{figure*}

In Figure~\ref{fig:qae_performance} we observed that inserting a PQC in a classical autoencoder enhances the anomaly detection performance. To establish that the performance-boost in the HAE cannot be solely attributed to the inherent ability of a PQC to embed data into a higher dimensional Hilbert space, we perform a small test. We replace the PQC part in HAE (see Figure~\ref{fig:hybrid_autoencoder}) with three classical layers, consisting of \texttt{4}, \texttt{16} and \texttt{4} nodes respectively. The new classical layers inflate the latent space from $4$ dimensions to $16$ (which is a size of the Hilbert space) dimensions and deflating back to the $4$ dimensions again. An architectural sketch of this new model is represented in Fig.~\ref{fig:newarchitecture}

In the next section, we compare the performance of this modified AE against AE and HAE models on a more realistic and complex task.

\section{Results}\label{sec:final_arch_hae_gas_turbine}

After examining the performances of the HAE with different PQC, we select the circuit design that exhibited the largest performance boost for further testing. Figure~\ref{fig:qae_performance} shows that Circuit 10 (see Figure~\ref{circuit_10}) provided the highest recall and F1 scores of the different circuits tested. For the remainder of the paper, we use Circuit 10 as the defining PQC of our HAE model.

\begin{figure}[htb]
    \centering
    \includegraphics[width=0.5\textwidth]{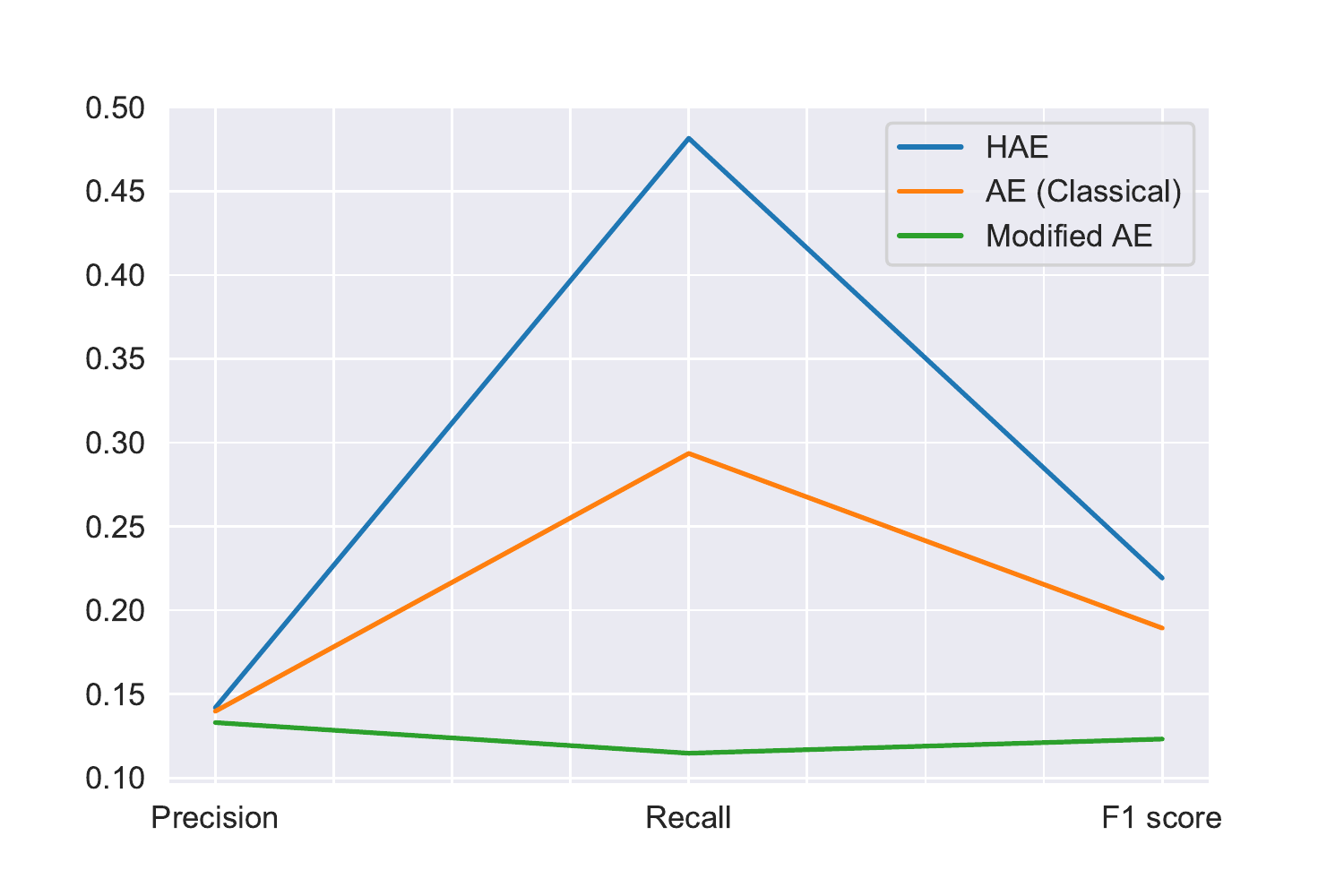}
    \caption{Precision, recall, and F1 score comparison of AE, modified AE (Section~\ref{section:quantum layer is replaced by classical layers}), and HAE with Circuit 10 (Figure~\ref{circuit_10}) tested on Gas turbine dataset with 2000 datapoints.}
    \label{fig:qVAE_versus_all_new}
\end{figure}

As a final application of the HAE model for the Gas turbine anomaly detection problem, we compare its performance in a more operationally realistic and difficult scenario. 
We first train the HAE on $1$ week of data (640 data points), use the encoder and PQC layers to obtain the dimension reduced representation of the data, and train a simple Isolation forest model on the latent representation of the training data. Then, we simulate the model running live in operations where we first embed each new data point from the test set into its dimension reduced form, which we then label as outlier or inliers using the trained Isolation forest. The results of this benchmark are shown in Figure \ref{fig:qVAE_versus_all_new} where we compare to the purely classical neural network architecture (AE), which contains the same encoder and decoder sizes, the latent space expansion model (Modified AE) described in Section~ \ref{section:quantum layer is replaced by classical layers}, as well as our proposed hybrid model (HAE). It can be seen that the HAE provided a performance enhancement in terms of precision, recall, and F1 score with respect to the anomaly class, in which we obtain the highest increase of $64.1 \%$ in recall. However, for the modified AE model, we see that all metric scores have been reduced as compared to the HAE model. In \cite{hilbert_feature} and \cite{schuld2021supervised}, the authors showed that the process of encoding inputs in a quantum state is a nonlinear feature map that maps data to a (classically intractably large) quantum Hilbert space. These ideas and the results of our small experiment contribute to the fact that just expanding the latent space is not enough to see observed benefits of inserting a PQC.

\subsection{Benchmarking with publicly available datasets}\label{section:Benchmarking with Standard datasets}

\begin{table}[htb]
    \centering
    \begin{tabular}{l|l|l|l|l}
          \textbf{Dataset}&\parbox{2.3cm}{\textbf{Performance}\\ \textbf{metric}}  & \textbf{AE}  & \textbf{HAE} & \parbox{2cm}{\textbf{Improvement}\\\textbf{Percentage}} \\
         \hline
         & Precision  & 0.855	 &  \textbf{0.882} & 3.1\\
         Musk & Recall		& 0.969	& \textbf{1.000} & 3.2\\
         &F1 score	 & 0.908	& \textbf{0.937} & 3.2\\
         \hline
         & Precision  & \textbf{0.662}	 &  0.623 & -5.8\\
         Arrythmia & Recall		& 0.652	& \textbf{0.727} & 11.5\\
         & F1 score	 & 0.656	& \textbf{0.671} & 2.3\\   
         \hline
         & Precision  & 0.523	 &  \textbf{0.613} & 17.2\\
         Satellite & Recall	 	& 0.680	& \textbf{0.760}& 11.7\\
         &F1 score	 & 0.591	& \textbf{0.679} & 14.9\\
    \end{tabular}
    \caption{Comparing AE to HAE on multiple public datasets. HAE provides an improvement in almost all categories for the three datasets (highlighted in bold).}
    \label{tab:summary_all_dataset}
\end{table}

In this section, we benchmark our algorithms with publicly available datasets as described in Section~\ref{subsection:additional_dataset}. These datasets are multi-class classification datasets, in which minority classes were merged together to represent an outlier class and the rest is considered to be healthy data.

From the above datasets we prepare three pairs of training and test datasets in which each training dataset consists of $320$ data points. We apply the same training and testing technique as described previously for the Gas turbine dataset. Table~\ref{tab:summary_all_dataset} summarizes the results of all AE and HAE performances on all three datasets. It can be seen that the hybrid classical-quantum model (HAE) is a better performing model in terms of not only recall, which had the highest improvement, but also precision and therefore F1 score on almost all datasets. The performance of models on these datasets is much more reliable than on the Gas turbine dataset due to the difficulties described in Section~\ref{subsection:benchmark}.

\section{Summary and Conclusion}\label{section:conclusion}

In this article, we have introduced a novel approach of an autoencoder hybridization that shows a performance increase for anomaly detection problem on quantum computers that are available today. We speculate that this can be attributed to the connection between quantum neural networks and kernel methods, as highlighted in  \cite{hilbert_feature, schuld2021supervised}, due to the nature of anomaly detection in our method. These models analyse the data in a high-dimensional Hilbert space, which is classically intractable and only accessible through the inner products and revealed by measurements. However, we leave the analysis of the deeper connection between our model and kernel methods to future work.

We probed over 30 different PQCs to find a suitable candidate for our HAE model, and we have performed an investigation into what makes some circuits more effective than the others. The evidence from this study suggests that the strength of the entanglement has a high impact on the model performance, which supports the ideas presented in \cite{circuit_centric} as the model is apt to capture correlation within the input data better. Our study provides a stimulus to an investigation of Fourier analysis-based expressivity measures for quantum models as proposed in \cite{encoding_fourier}, as our results indicate that Fourier inspired descriptors have positive correlations with the models' performance in terms of precision, recall and F1 score. The authors of the original paper provided us with tools to control the amount of frequencies in the spectrum (by repeating the embedding layer), however, the control of Fourier coefficients' expressivity is still an open question, which can be addressed in further research

The study has also shown that circuits which are more expressive in terms of Sim expressivity \cite{expressibility_haar} have smaller reconstruction loss, which aligns with the results from \cite{similar_paper}. This was the only quantum circuit metric that assisted during the training of a model. The authors of \cite{funcke2021dimensional} give insight into how to design PQCs with maximal expressivity and minimal amount of trainable parameters.

Our model has displayed a peculiar behaviour: the more complex circuits showed higher reconstruction loss and at the same time better ability to identify anomalies. Another oddity is that our experiment did not reveal a strong correlation between precision, recall or F1 score with Fourier spectrum size, which is responsible for the complexity of the functions that our PQC can represent. One possible reason for these behaviours, as mentioned before, is that we do not adjust the hyperparameters for training progressively complexer models. The reason we kept these parameters static is to allow for fair comparison between different PQCs. However, one can explore more adaptable training schemes to accommodate for this issue. Another possible explanation for these phenomena is that our training approach does not foresee encountering barren plateaus (e.g. \cite{McClean_2018, barren_identity, barren_layerwise, Cerezo_2021}). We suspect that our chosen circuits are shallow enough to avoid this issue, however, this point could be investigated further.

One further intriguing question remains open: how does different measurement strategies affect model performance? Currently, as described in Section~\ref{section:hybrid_autoencoder}, we calculate expectation values for each qubit separately and by doing so loose correlation information. An interesting augmentation of this experiment would be to test different measurement and measurement post-processing strategies and see which ones are more beneficial for this task and why.

We believe that this study has shown a promising hybrid approach for anomaly detection research direction, as well as some insights into which features of PQCs might be beneficial for this task. This work opens a revenue for further hybrid autoencoder based models, like a hybrid version of a variational autoencoder for gate-based computers that provides an additional increase in performance and is the topic of our current research. This work provides a springboard for further research into different aspects of PQCs that can be leveraged to increase the performance of classical models, as well as Fourier-analysis based expressivity measures.



\bibliography{ref}



\appendix{
\section{Data preprocessing step for Gas turbine dataset}\label{section:data_proprocessing}
Value ranges of different sensors are highly heterogeneous. Therefore, as a first preprocessing step, the data is rescaled to fit in the interval $[0,1]$. Once the data is rescaled, it has to be denoised. This is one of the most fundamental steps in data preprocessing, which is especially crucial when working with real data. First, we need to identify what healthy data is, to see which data points do not fit into the pattern. Clustering is an effective method to achieve that. One of the prominent clustering methods is \textit{Density-Based Spatial Clustering of Applications with Noise (DBSCAN)}, which assigns high-density areas to a cluster and marks data instances that are too far away as anomalies. In this project, we rely on \texttt{sklearn.cluster.DBSCAN} \cite{pedregosa2011scikit} implementation of this algorithm. The algorithm's performance heavily relies on the choice of parameters, such as \texttt{eps} and \texttt{min\_samples}:
\begin{itemize}
    \item \texttt{eps} controls the epsilon distance around a point, within which all other points will be assigned to its neighbourhood. To find a suitable value of this parameter, we can use a method that is known in the machine learning folklore as an \textit{elbow method} and was formally introduced by \textit{Rahmah et al.}~\cite{Rahmah_2016}. The idea behind this technique is to compute all the pair-wise distances between points of the dataset, sort them in ascending order and plot them. The plot will reveal the point in which there is a drastic increase in distances. This rapid change indicates the rarity of those distances in the dataset.
    \item \texttt{min\_samples} parameter controls the minimal size of the neighbourhood for a point to be considered a core. Here, it is set to $2\%$ of the training dataset size. 
\end{itemize}
\texttt{sklearn.cluster.DBSCAN} assigns a label for each data instance in the dataset that indicates which cluster the point was allocated to. Noise samples are labelled $-1$, which are the ones that we can remove.

\section{List of Quantum Circuits}\label{section:circuit_list}
\begin{figure}[htb]
     \centering
    \begin{subfigure}[b]{0.2\textwidth}
         \centering
         \begin{quantikz}
         \lstick{$\ket{0}$} & \gate{R_x} &\qw &\qw \\
         \lstick{$\ket{0}$} & \gate{R_x} &\qw &\qw \\
    \end{quantikz}
    \caption{Pauli-$X$ embedding}
     \end{subfigure}~
    \begin{subfigure}[b]{0.2\textwidth}
         \centering
         \begin{quantikz}
         \lstick{$\ket{0}$} & \gate{R_y} &\qw &\qw \\
         \lstick{$\ket{0}$} & \gate{R_y} &\qw &\qw \\
    \end{quantikz}
    \caption{Pauli-$Y$ embedding}
     \end{subfigure}
     
    \begin{subfigure}[b]{0.2\textwidth}
         \centering
         \begin{quantikz}
         \lstick{$\ket{0}$} & \ctrl{1} &\qw &\qw \\
         \lstick{$\ket{0}$} & \gate{CR_x} &\qw &\qw \\
    \end{quantikz}
    \caption{Controlled Pauli-$X$ embedding}
     \end{subfigure}~
    \begin{subfigure}[b]{0.2\textwidth}
         \centering
         \begin{quantikz}
         \lstick{$\ket{0}$} & \gate{H} & \gate{R_z} &\qw \\
         \lstick{$\ket{0}$} & \gate{H} & \gate{R_z} &\qw \\
    \end{quantikz}
    \caption{Hadamard Pauli-$Z$ embedding}
     \end{subfigure}

     \caption{Different types of gate-based embedding layers that were used in this project. These layers can be inserted into a PQC one or multiple times.}
     \label{fig:embeddings}
\end{figure}
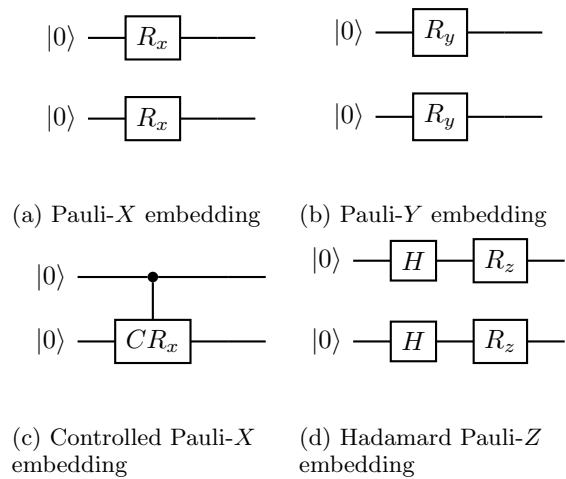

We have tested 32 parametrized quantum circuits in this project with 4 different embedding strategies: Pauli-$X$ (Figures~\ref{circuits_1_9}-\ref{circuits_14_21}), Pauli-$Y$ (Figures~\ref{circuits_14_21}-\ref{circuits_22_26}), Controlled Pauli-$X$ (Figures~\ref{circuits_22_26}-\ref{circuits_27_32}) and Hadamard-Pauli-$Z$ (Figure~\ref{circuits_27_32}) and their redundancies (repeated embedding layers). All embedding layers are highlighted with \texttt{E}, all quantum gate notations follow convention except for $R_*(x)$ gates, which indicate gates that with rotation axes chosen at random.

Circuit 7 (Figure~\ref{circuit_7}), Circuit 15 (Figure~\ref{circuit_15}), Circuit 18 (Figure~\ref{circuit_18}) and Circuit 25 (Figure~\ref{circuit_25}) exhibited a high reconstruction loss and therefore failed to converge during training. The statistics from these circuits were excluded from overall consideration.

\begin{figure*}
    \captionsetup[subfigure]{labelformat=empty}
    \subcaptionbox{Circuit 1\label{circuit_1}}
        {\includegraphics[scale=.7]{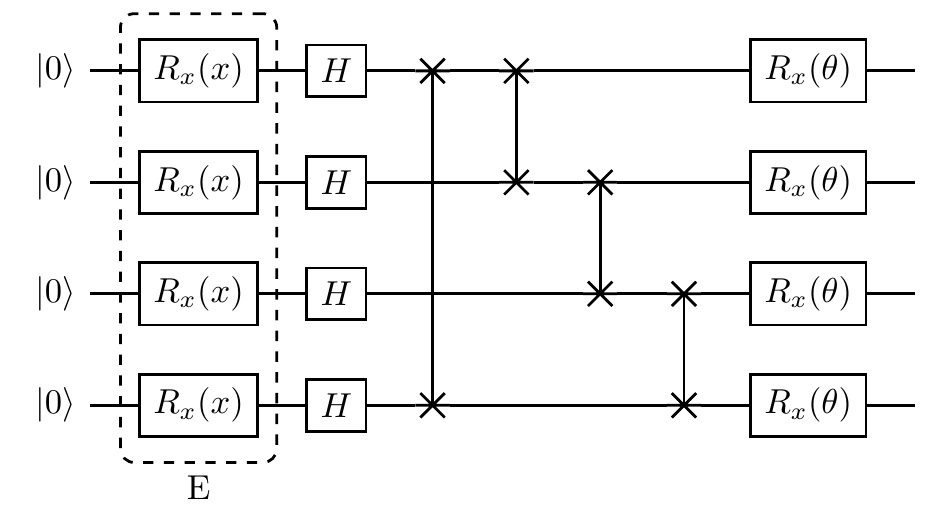}}
    \subcaptionbox{Circuit 2\label{circuit_2}}
        {\includegraphics[scale=.7]{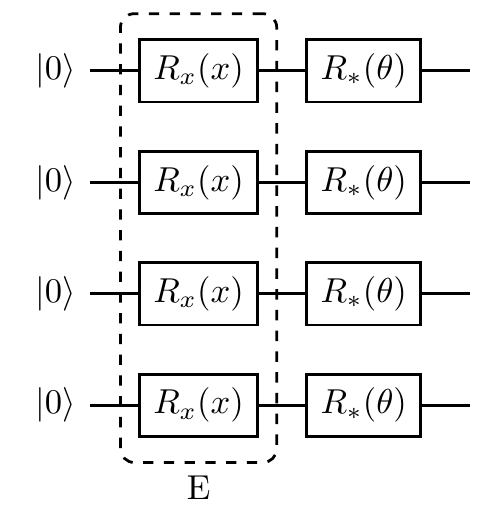}}
    \subcaptionbox{Circuit 3\label{circuit_3}}
        {\includegraphics[scale=.7]{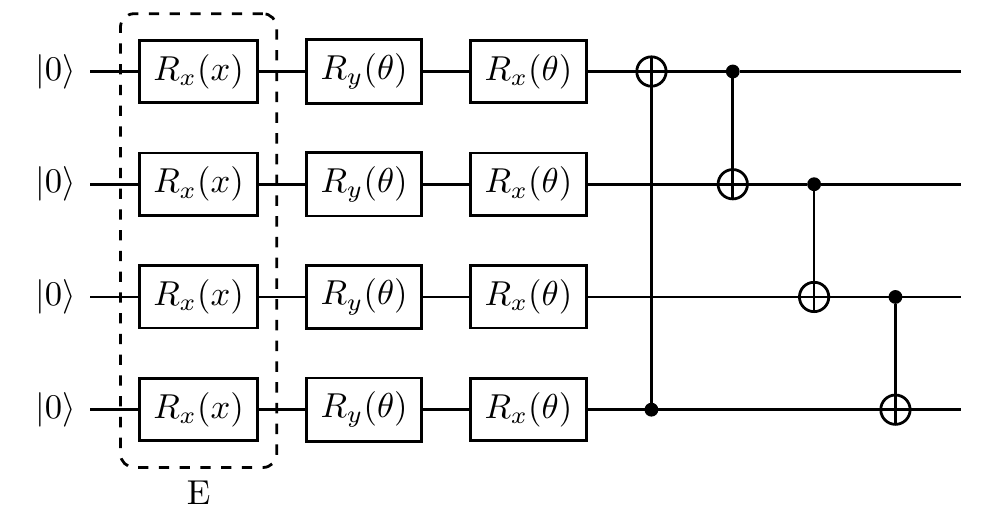}}
    \subcaptionbox{Circuit 4\label{circuit_4}}
        {\includegraphics[scale=.7]{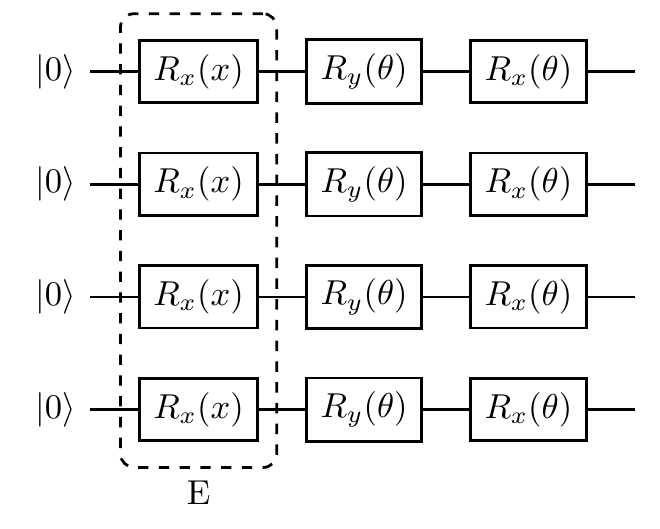}}
    \subcaptionbox{Circuit 5\label{circuit_5}}
        {\includegraphics[scale=.7]{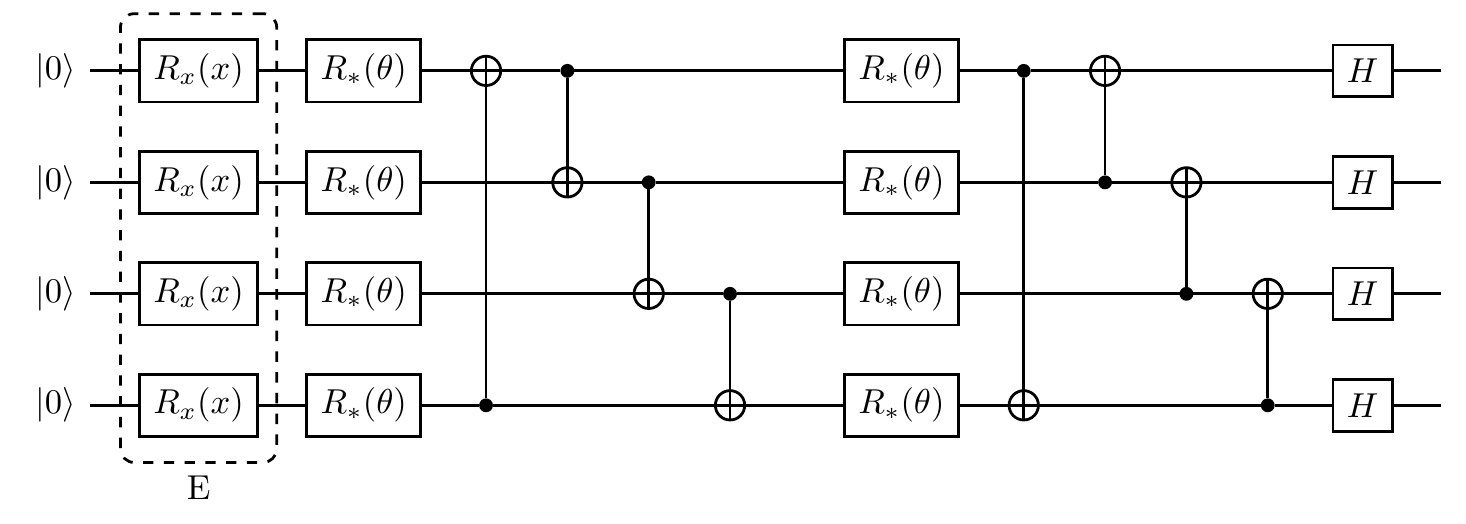}}
    \subcaptionbox{Circuit 6\label{circuit_6}}
        {\includegraphics[scale=.7]{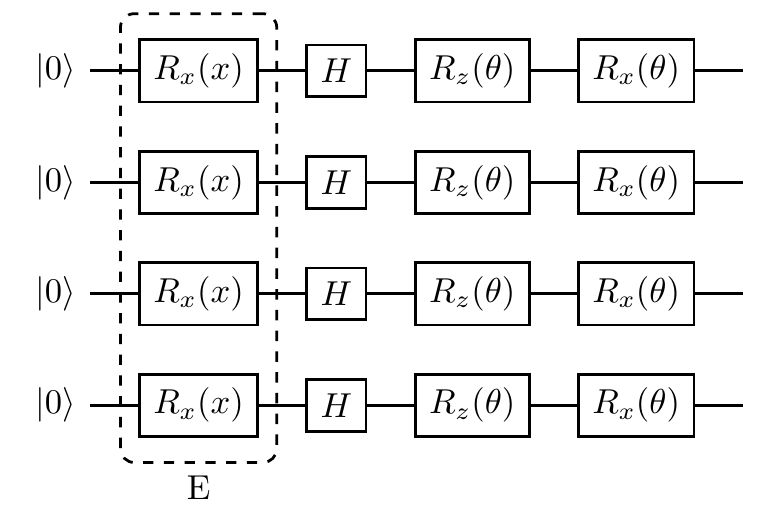}}
    \subcaptionbox{Circuit 7\label{circuit_7}}
        {\includegraphics[scale=.7]{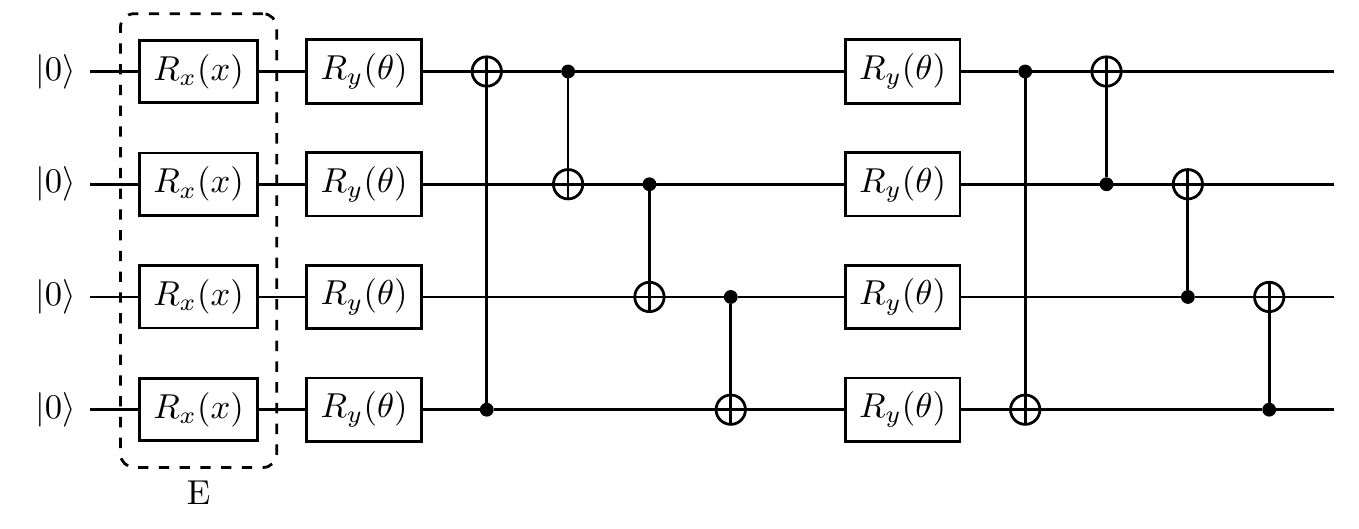}}
    \subcaptionbox{Circuit 8\label{circuit_8}}
        {\includegraphics[scale=.7]{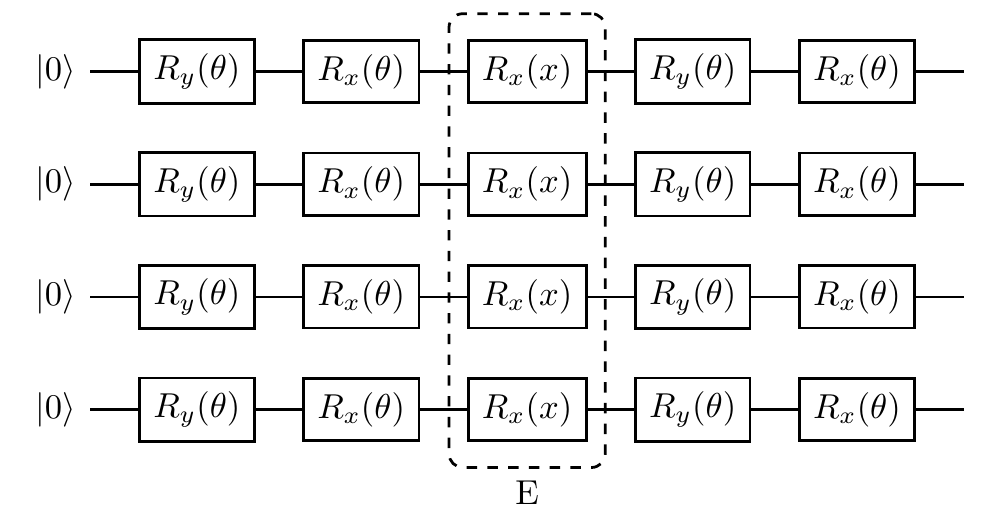}}
    \subcaptionbox{Circuit 9\label{circuit_9}}
        {\includegraphics[scale=.7]{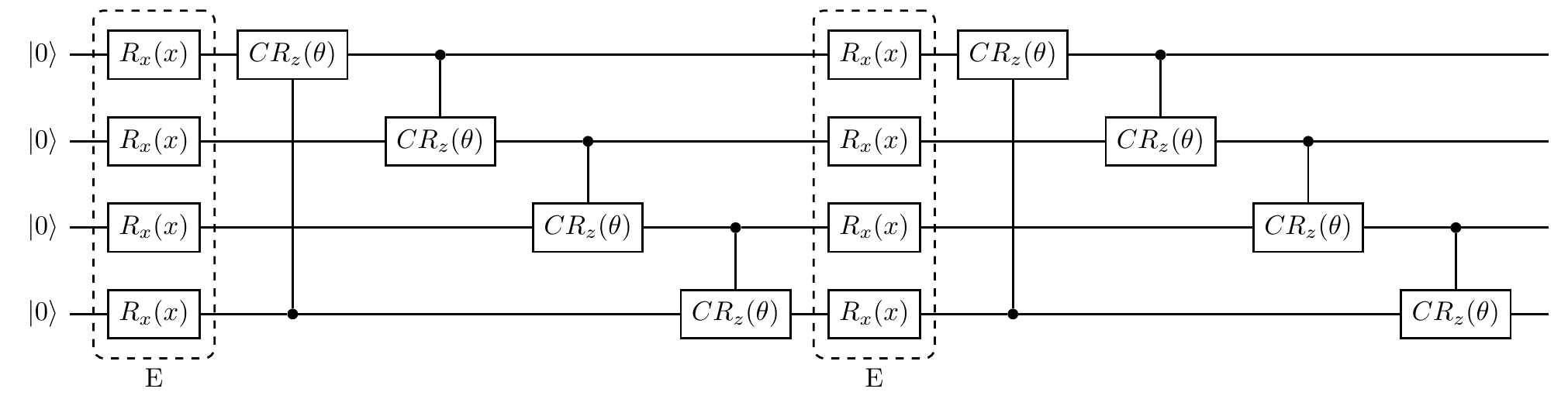}}
    \caption{List of quantum circuits with IDs 1 - 9}\label{circuits_1_9}
\end{figure*}

\begin{figure*}
    \captionsetup[subfigure]{labelformat=empty}
    \subcaptionbox{Circuit 10\label{circuit_10}}
        {\includegraphics[scale=0.7]{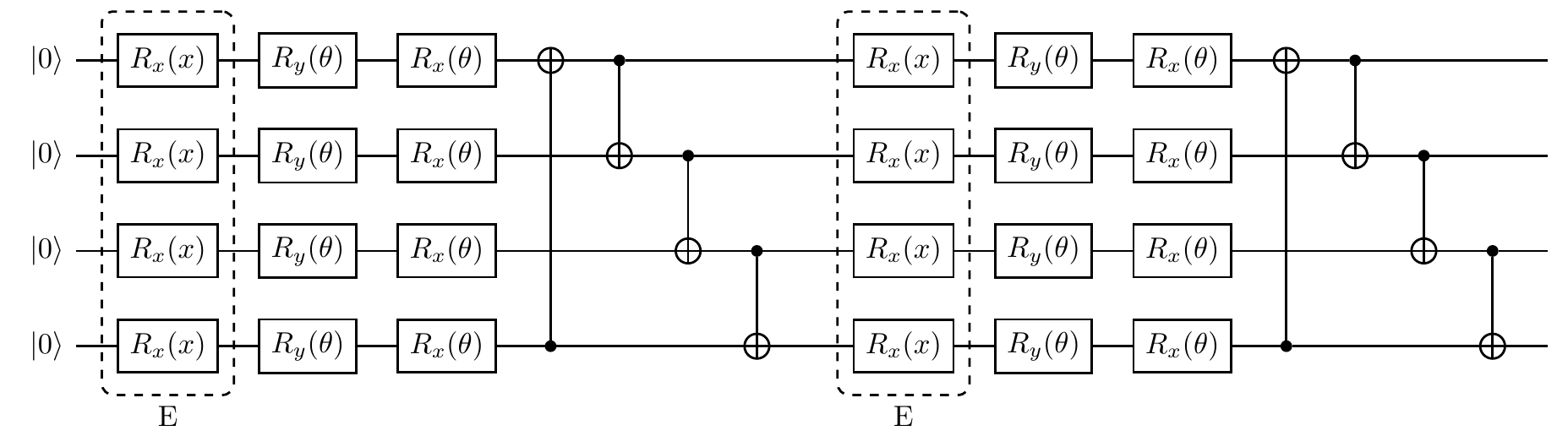}}
    \subcaptionbox{Circuit 11\label{circuit_11}}
        {\includegraphics[scale=0.7]{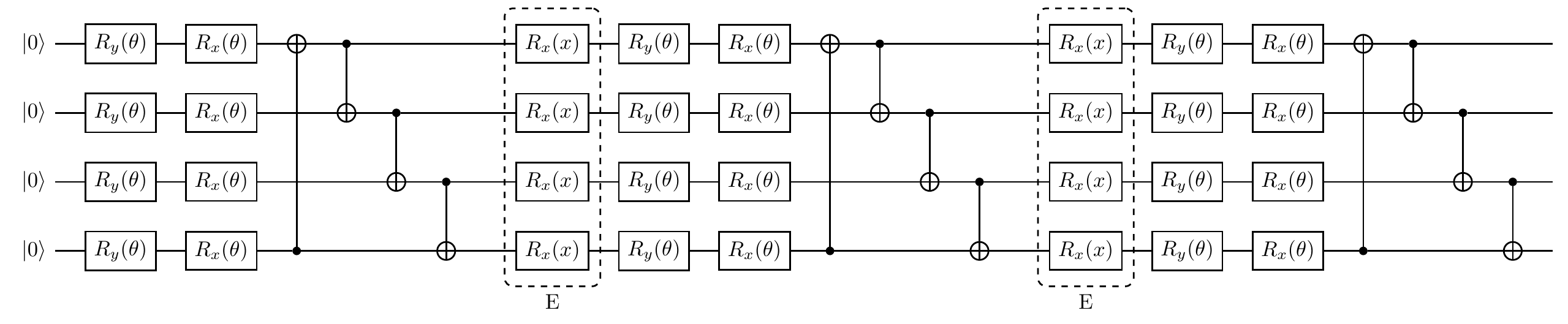}}
    \subcaptionbox{Circuit 12\label{circuit_12}}
        {\includegraphics[scale=0.7]{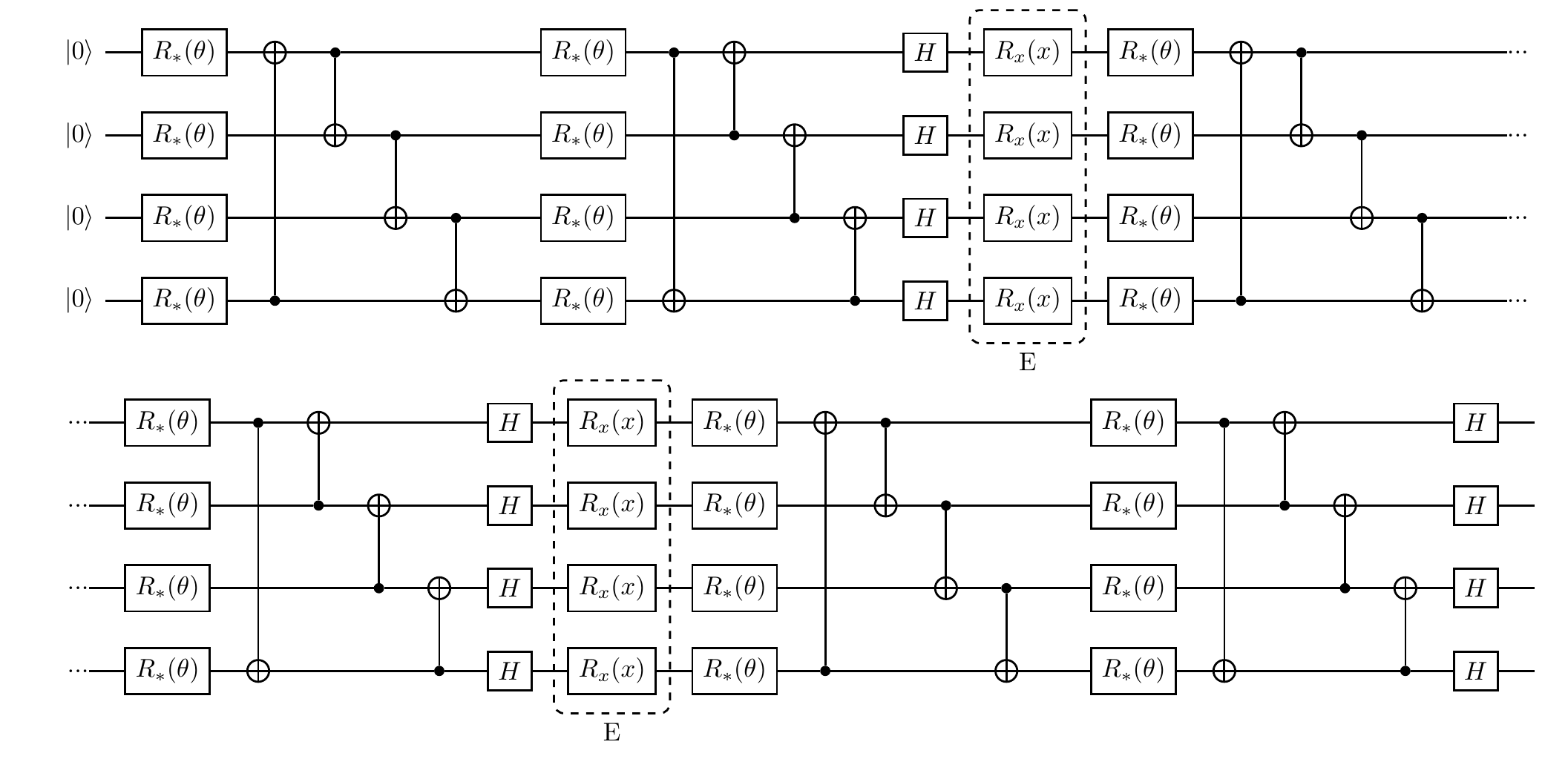}}
    \subcaptionbox{Circuit 13\label{circuit_13}}
        {\includegraphics[scale=0.7]{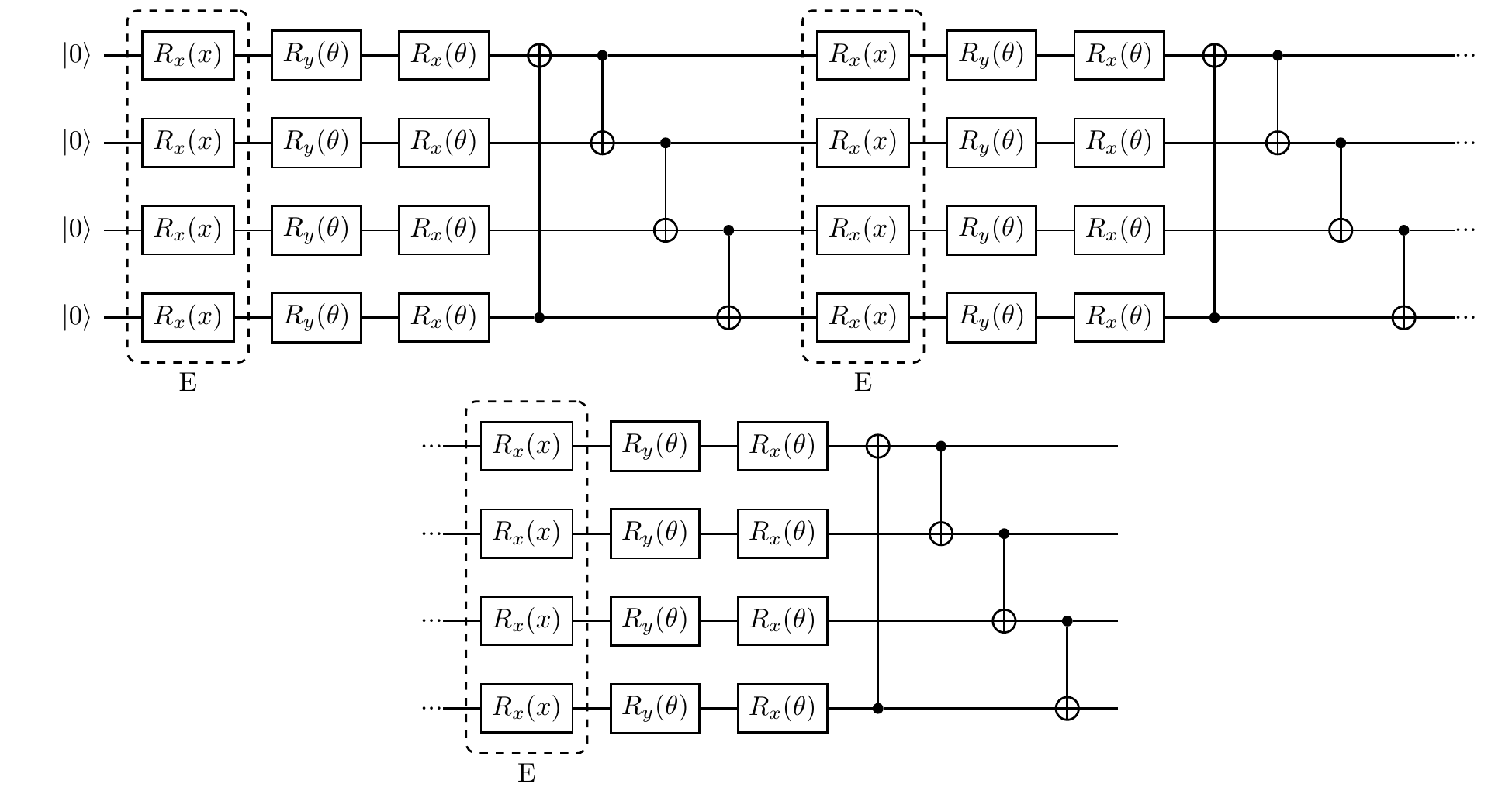}}
    \caption{List of quantum circuits with IDs 10 - 13}\label{circuits_10_13}
\end{figure*}

\begin{figure*}
    \captionsetup[subfigure]{labelformat=empty}
    \subcaptionbox{Circuit 14\label{circuit_14}}
        {\includegraphics[scale=0.7]{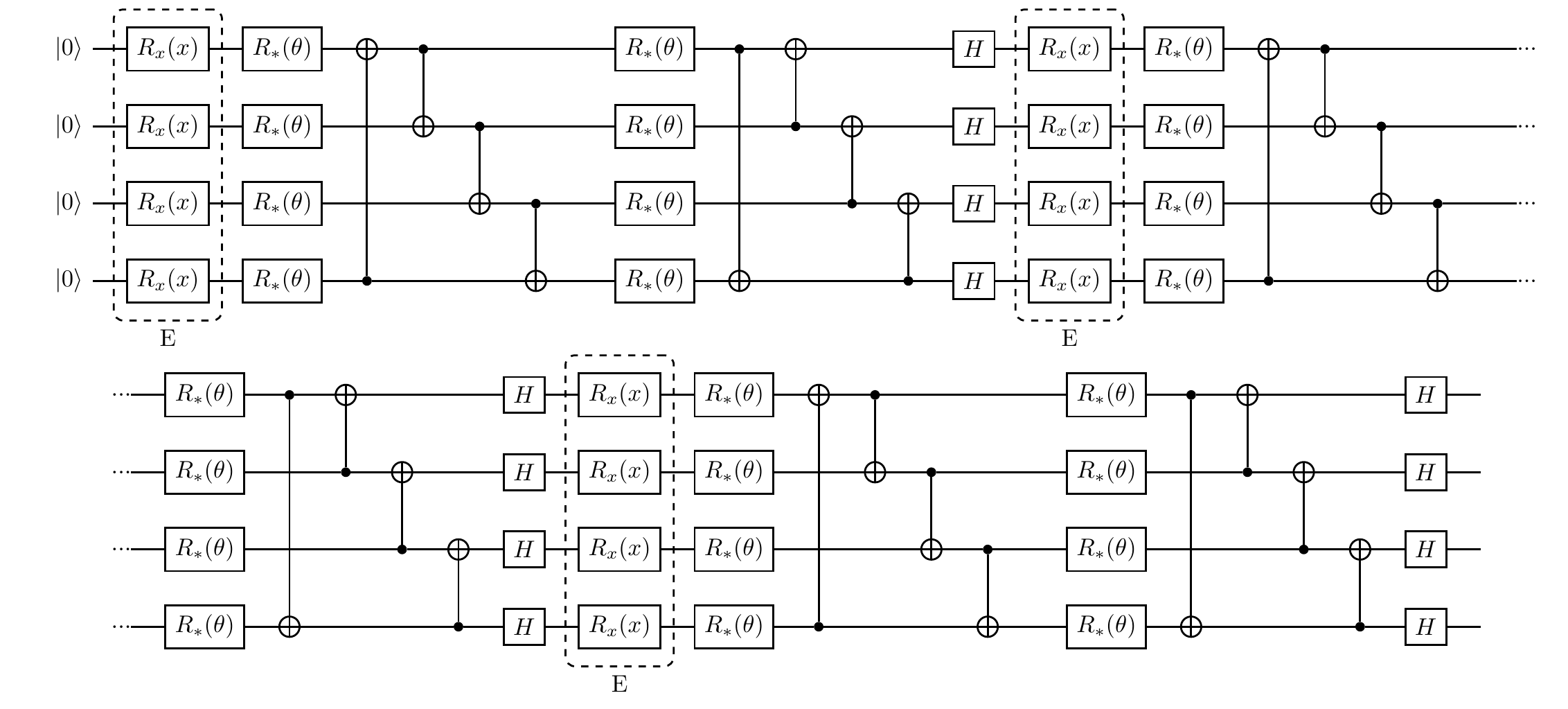}}
    \subcaptionbox{Circuit 15\label{circuit_15}}
        {\includegraphics[scale=0.7]{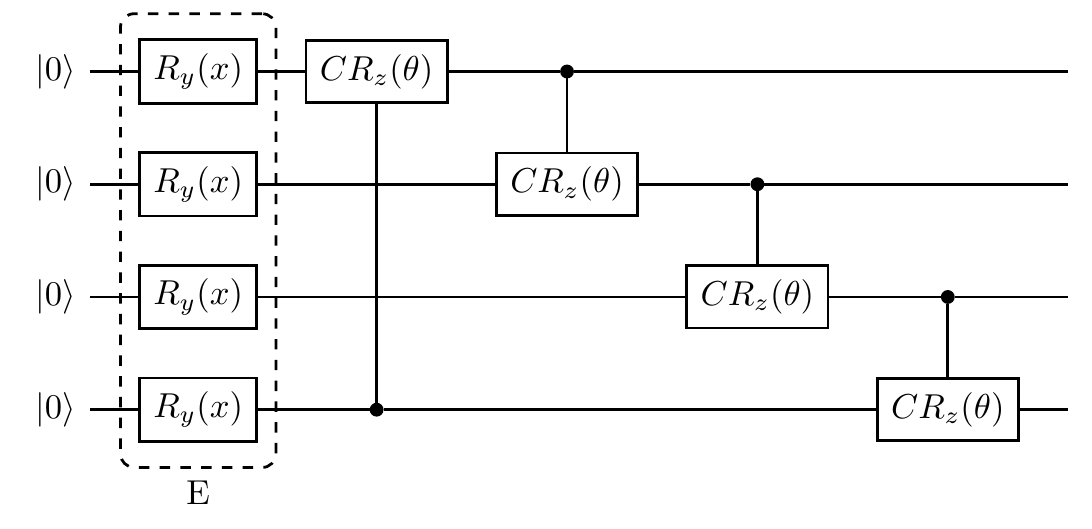}}
    \subcaptionbox{Circuit 16\label{circuit_16}}
        {\includegraphics[scale=0.7]{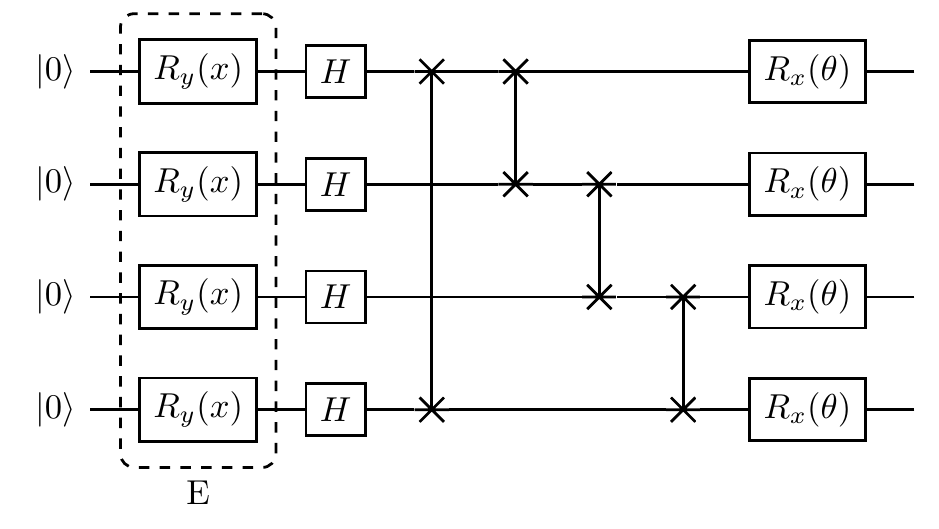}}
    \subcaptionbox{Circuit 17\label{circuit_17}}
        {\includegraphics[scale=0.7]{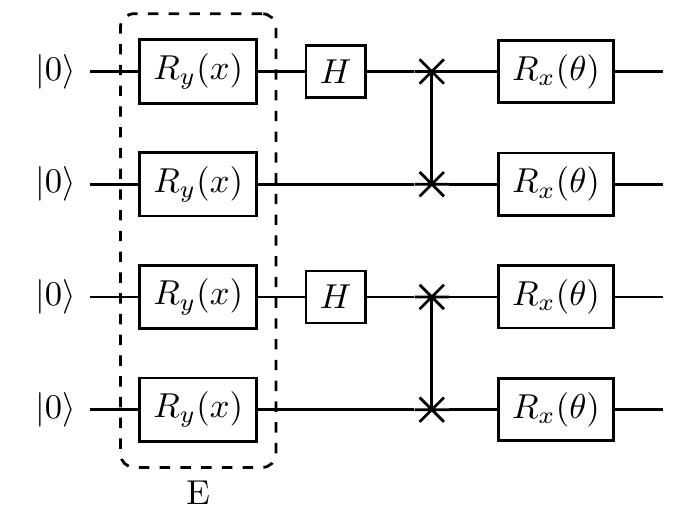}}
    \subcaptionbox{Circuit 18\label{circuit_18}}
        {\includegraphics[scale=0.7]{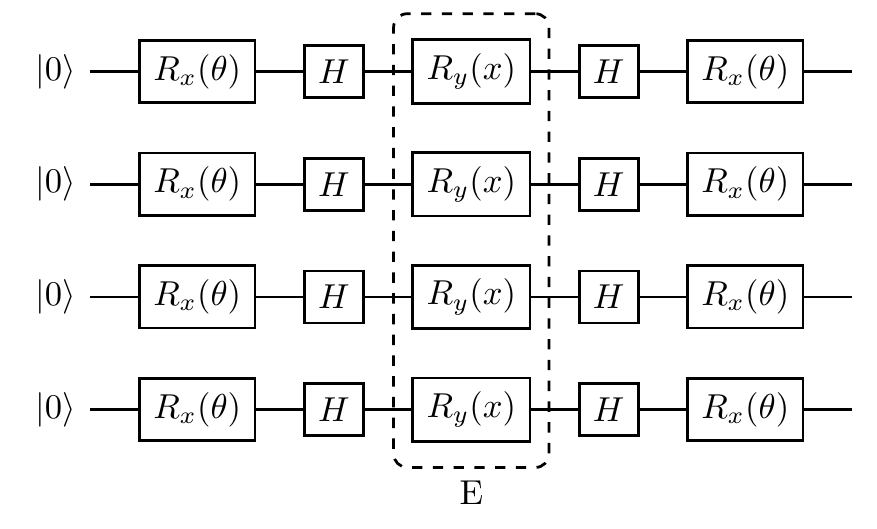}}
    \subcaptionbox{Circuit 19\label{circuit_19}}
        {\includegraphics[scale=0.7]{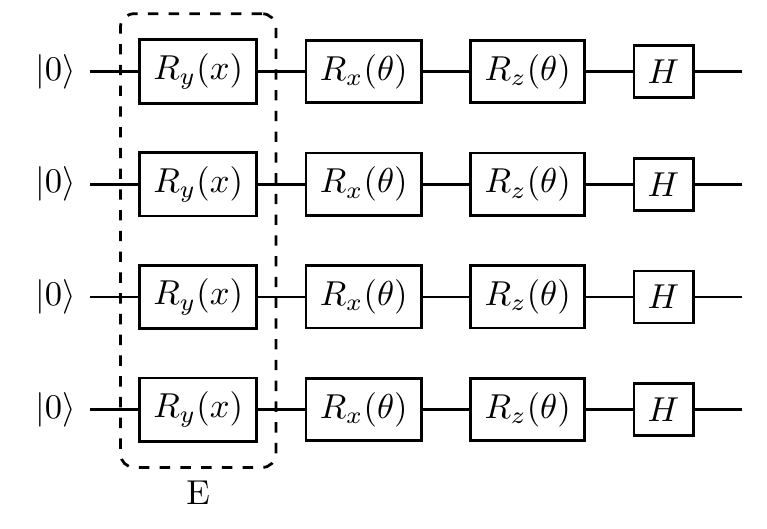}}
    \subcaptionbox{Circuit 20\label{circuit_20}}
        {\includegraphics[scale=0.7]{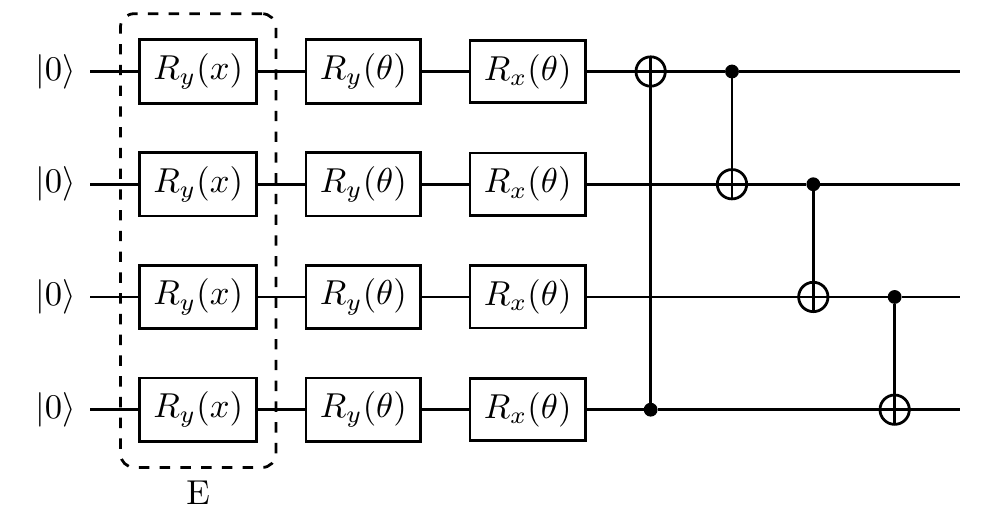}}
    \subcaptionbox{Circuit 21\label{circuit_21}}
        {\includegraphics[scale=0.7]{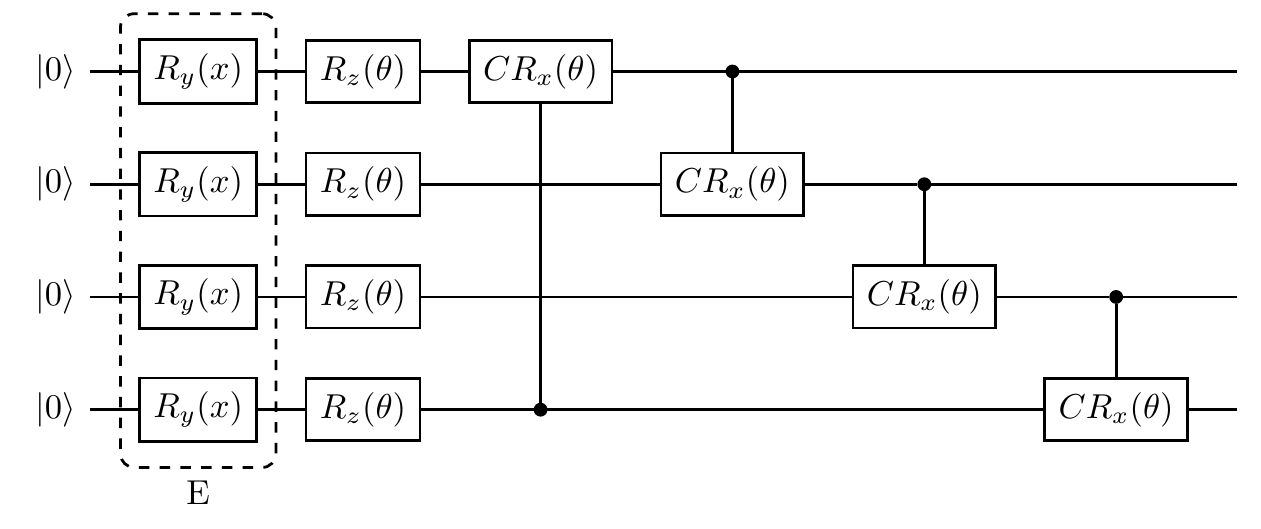}}
    \caption{List of quantum circuits with IDs 14 - 21}\label{circuits_14_21}
\end{figure*}

\begin{figure*}
    \captionsetup[subfigure]{labelformat=empty}
    \subcaptionbox{Circuit 22\label{circuit_22}}
        {\includegraphics[scale=0.7]{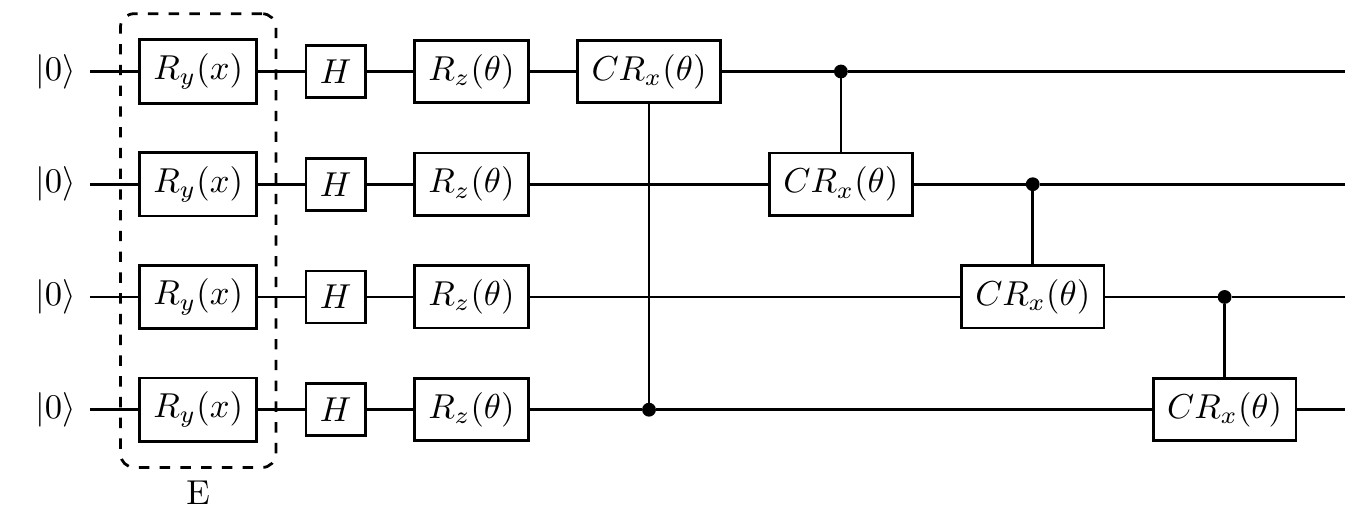}}
    \subcaptionbox{Circuit 23\label{circuit_23}}
        {\includegraphics[scale=0.7]{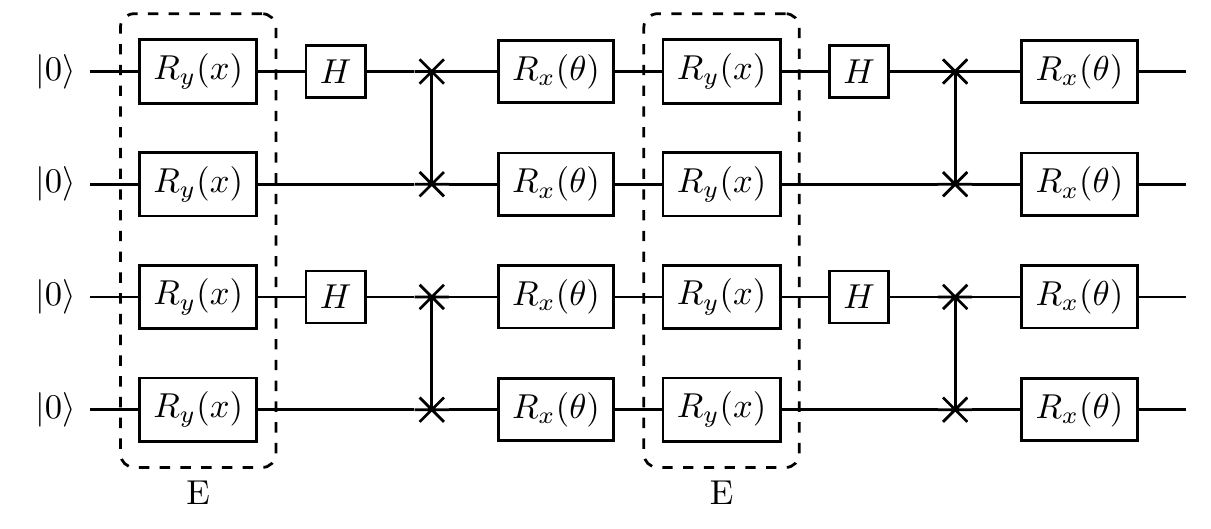}}
    \subcaptionbox{Circuit 24\label{circuit_24}}
        {\includegraphics[scale=0.7]{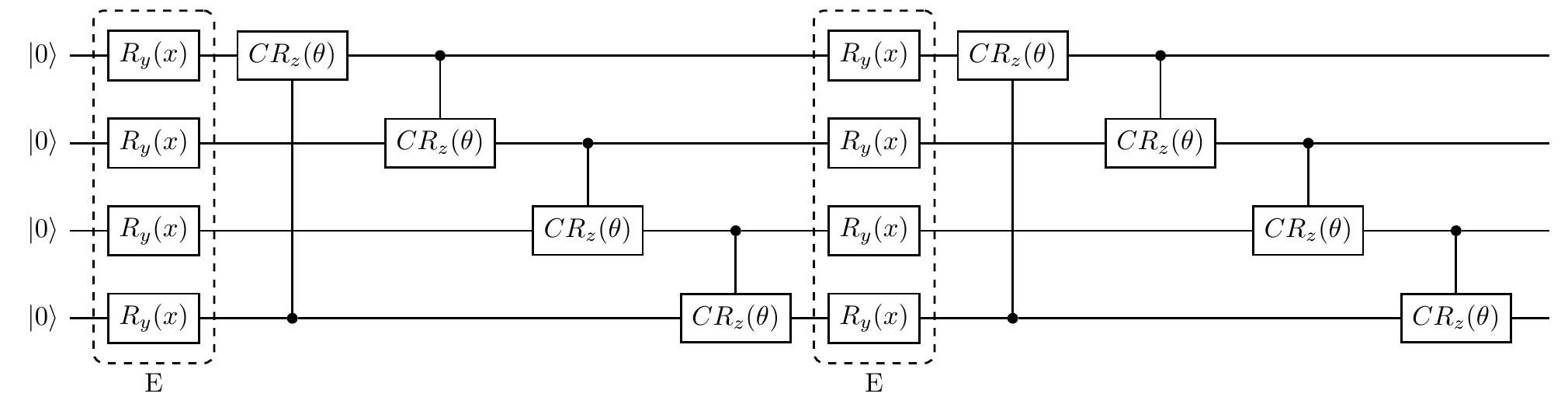}}
    \subcaptionbox{Circuit 25\label{circuit_25}}
        {\includegraphics[scale=0.7]{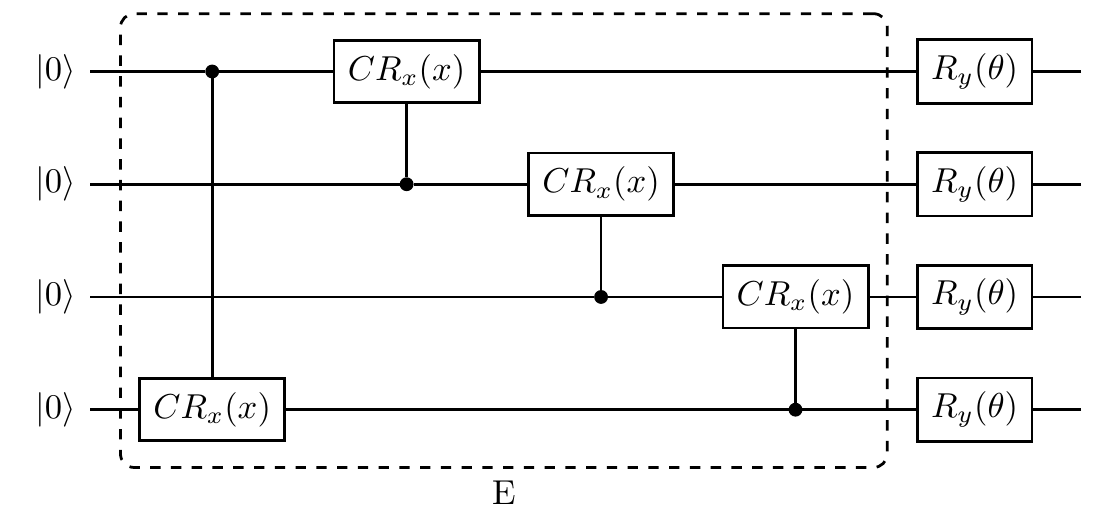}}
    \subcaptionbox{Circuit 26\label{circuit_26}}
        {\includegraphics[scale=0.7]{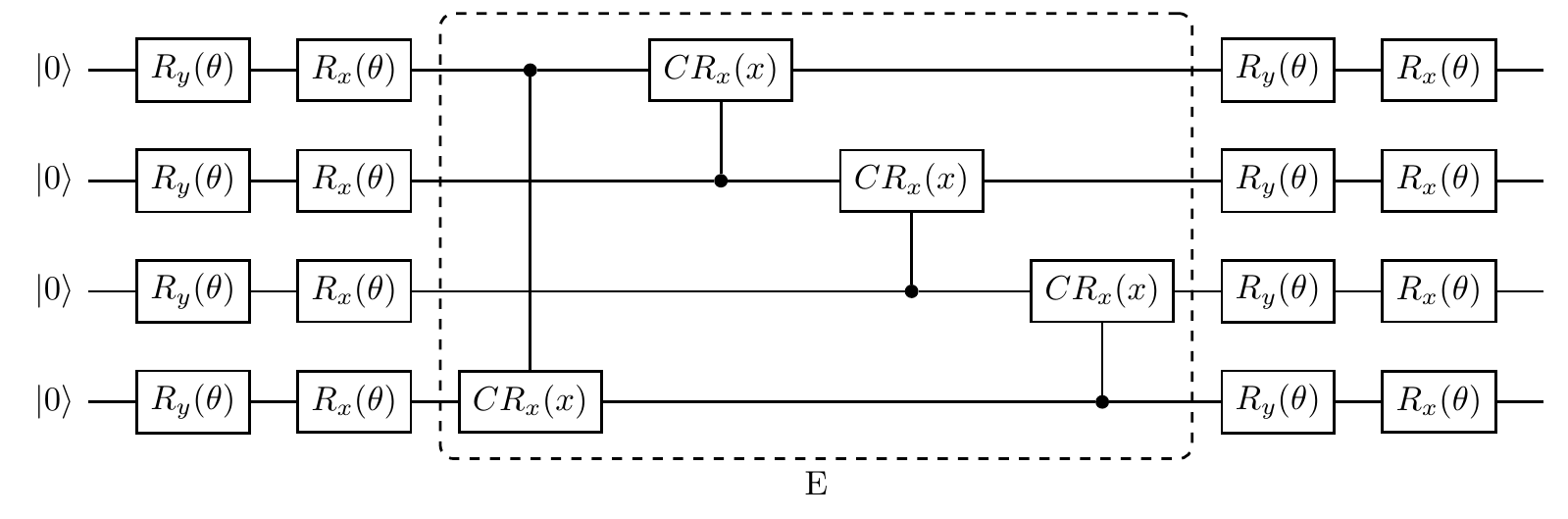}}
    \caption{List of quantum circuits with IDs 22 - 26}\label{circuits_22_26}
\end{figure*}

\begin{figure*}
    \captionsetup[subfigure]{labelformat=empty}
    \subcaptionbox{Circuit 27\label{circuit_27}}
        {\includegraphics[scale=0.7]{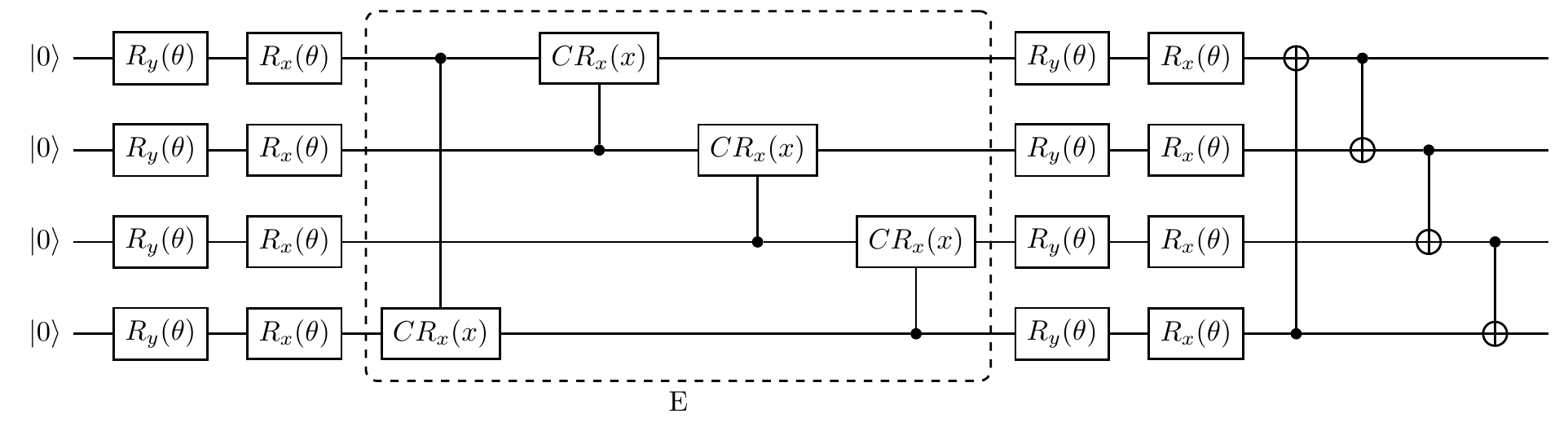}}
    \subcaptionbox{Circuit 28\label{circuit_28}}
        {\includegraphics[scale=0.7]{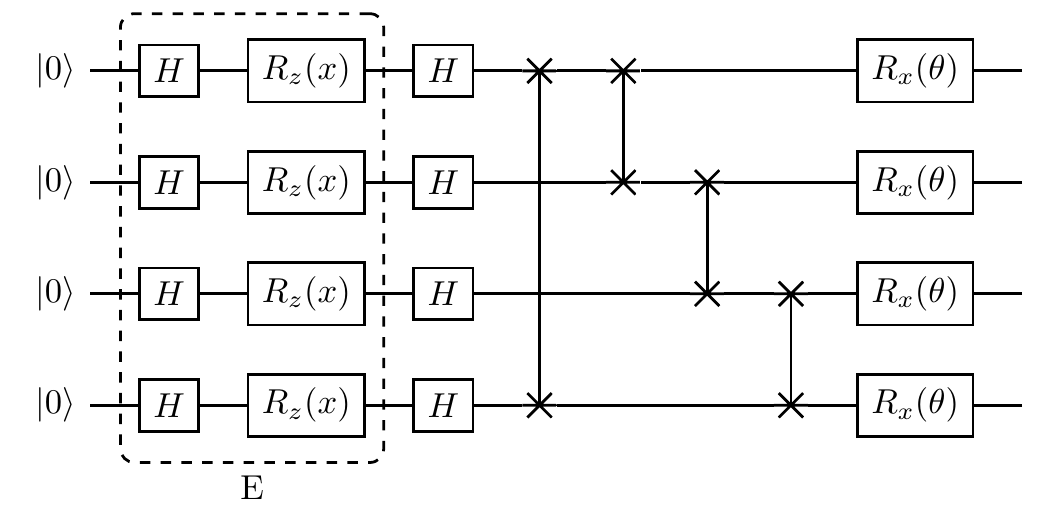}}
    \subcaptionbox{Circuit 29\label{circuit_29}}
        {\includegraphics[scale=0.7]{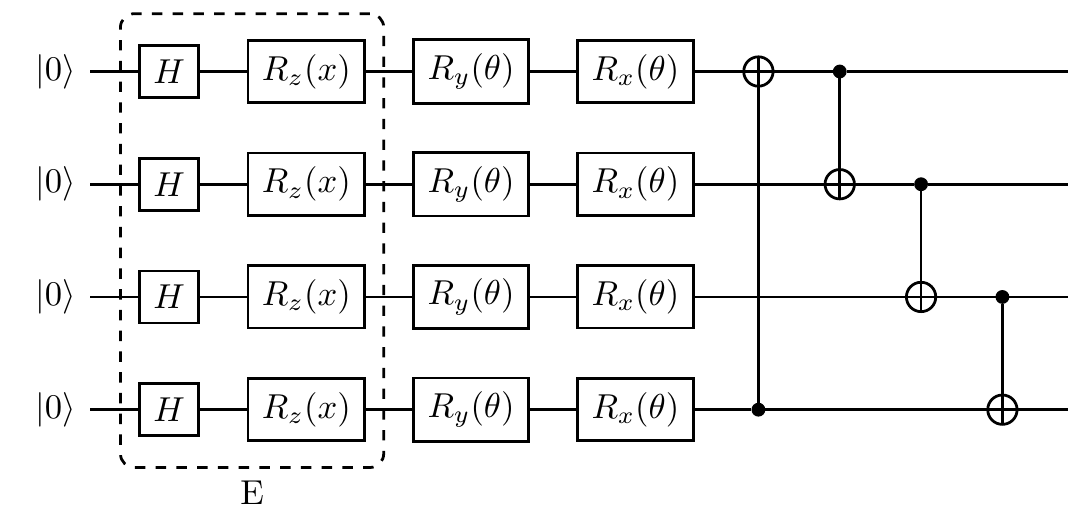}}
    \subcaptionbox{Circuit 30\label{circuit_30}}
        {\includegraphics[scale=0.7]{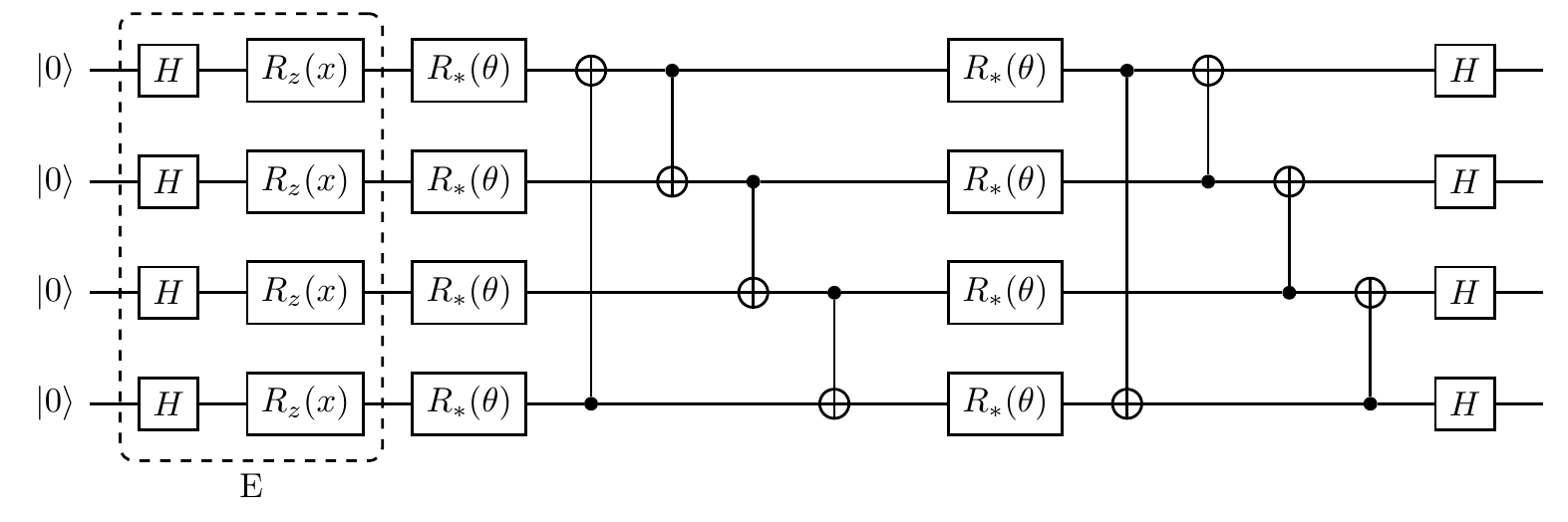}}
    \subcaptionbox{Circuit 31\label{circuit_31}}
        {\includegraphics[scale=0.7]{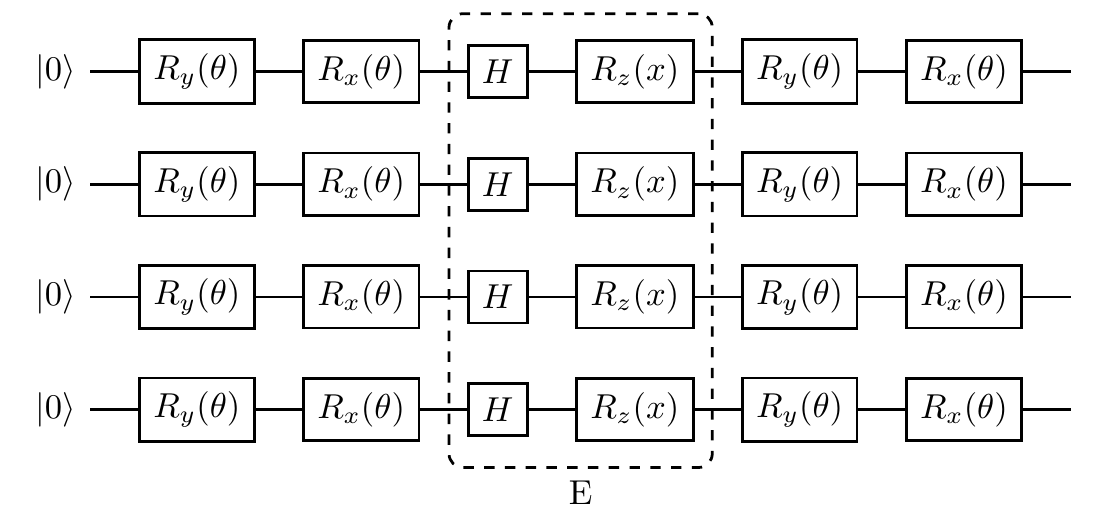}}
    \subcaptionbox{Circuit 32\label{circuit_32}}
        {\includegraphics[scale=0.7]{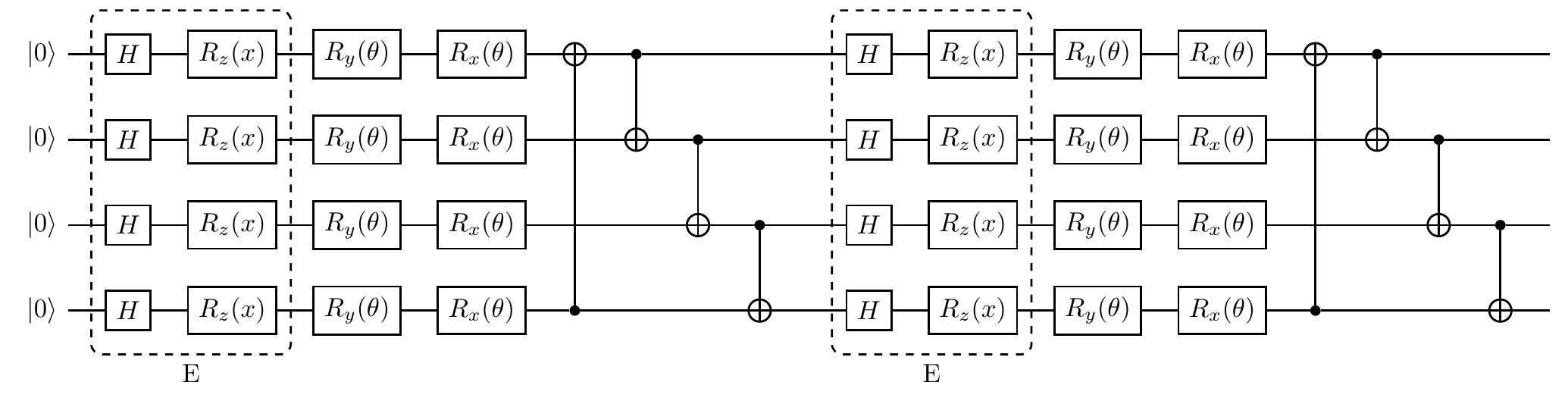}}
    \caption{List of quantum circuits with IDs 27 - 32}\label{circuits_27_32}
\end{figure*}

\section{Fourier Analysis of 1-Qubit System}~\label{appendix_fourier}
In order to gain a better insight into ideas presented in \cite{encoding_fourier}, we performed an experiment with a 1-qubit system. We concentrated on a single embedding technique (Pauli-$Y$) and varied its redundancy, its position in a quantum circuit and the trainable gates that surround it. The parameters $\theta$ of the trainable gates were selected at random $\theta \sim U[-\pi, \pi]$. The results of this experiment are presented in Table~\ref{tab:fourier_beauty}. 

The authors of \cite{encoding_fourier} have shown that a univariate quantum model $f$ can be represented through a partial Fourier series:
\begin{equation}\label{eq:fourier_series_exp}
    f(x) = \sum_{\omega \in \Omega} c_{\omega} e^{i \omega x},
\end{equation}
where $\Omega$ represents a frequency spectrum and $c_{\omega}$ stands for a coefficient that corresponds to a frequency $\omega$. The embedding redundancy influences the frequency spectrum $\Omega$ (Table~\ref{tab:fourier_beauty} displays only the positive spectrum) and hence impacts the complexity of the functions that a PQC can learn. For example, PQCs with a single Pauli-Y embedding has a single (positive) frequency at its disposal and hence can learn to represent only a cosine function. Increasing the redundancy of the embedding, increases the size of the spectrum and allows us to represent more intricate functions.

The way we construct the residual circuit impacts the amplitude and phase of the function as can be seen in Table~\ref{tab:fourier_beauty}. According to \cite{encoding_fourier} the design of the remainder gates affect the coefficients $c_{\omega}$. The connection between coefficients and Fourier phases and amplitudes becomes apparent once Eq.~\eqref{eq:fourier_series_exp} is represented in the amplitude-phase form:
\begin{equation}\label{eq:fourier_series_phase}
    f(x) = \frac{A_0}{2} + \sum_{\omega \in \Omega^+} A_{\omega} \cos(\omega x - \varphi_{\omega}),
\end{equation} 
where $\Omega^+$ is the positive spectrum and $\varphi_{\omega}$ is the phase of a frequency $\omega$. During the transition from Eq.~\eqref{eq:fourier_series_phase} to Eq.~\eqref{eq:fourier_series_exp}, 
$\Omega^+$ needs to be extended to a full $\Omega$, which can be done by utilizing Euler's formula:
\begin{align}
    \cos(\omega x - \varphi_{\omega}) & = \frac{1}{2}(e^{i(\omega x - \varphi_{\omega})} + e^{-i(\omega x - \varphi_{\omega})}) \nonumber\\
    & = \underbrace{\frac{1}{2}e^{-i \varphi_{\omega}}e^{i\omega x}}_{\Omega^+} + \underbrace{\frac{1}{2}\overline{e^{-i \varphi_{\omega}}}e^{i(-\omega) x}}_{\Omega^-},
\end{align}
where $\Omega^- \cup \{0\} \cup \Omega^+ = \Omega$. Therefore coefficients $c_{\omega}$ take the following form:
\begin{equation}
c_{\omega}=
    \begin{cases}
    A_0/2       & \quad \omega \in \{0\}\\
    A_{\omega}e^{-i \varphi_{\omega}}/2& \quad \omega \in \Omega^+\\
    A_{\omega}\overline{e^{-i \varphi_{\omega}}}/2 & \quad \omega \in \Omega^-,
  \end{cases}
\end{equation}
from where we see the dependence of coefficients $c_{\omega}$ on amplitudes $A_{\omega}$ and phases $\varphi_{\omega}$. 

\sbox0{\resizebox{!}{4em}{\begin{quantikz}[baseline={([yshift={-7\ht\strutbox}]current bounding box.north)}] \lstick{$\ket{0}$} & \gate{R_y(x)}\gategroup[1,steps=1,style={dashed, rounded corners,inner xsep=2pt}, background,label style={label position=below,anchor=north,yshift=-0.2cm}]{{E}} & \qw \end{quantikz}}}

\sbox1{\resizebox{!}{4em}{\begin{quantikz}[baseline={([yshift={-7\ht\strutbox}]current bounding box.north)}] \lstick{$\ket{0}$} & \gate{R_y(x)}\gategroup[1,steps=1,style={dashed, rounded corners,inner xsep=2pt}, background,label style={label position=below,anchor=north,yshift=-0.2cm}]{{E}} & \gate{R_x(\theta)} & \qw \end{quantikz}}}

\sbox2{\resizebox{!}{4em}{\begin{quantikz}[baseline={([yshift={-7\ht\strutbox}]current bounding box.north)}] \lstick{$\ket{0}$} & \gate{R_y(x)}\gategroup[1,steps=1,style={dashed, rounded corners,inner xsep=2pt}, background,label style={label position=below,anchor=north,yshift=-0.2cm}]{{E}} & \gate{H} & \gate{R_x(\theta)} & \qw \end{quantikz}}}

\sbox3{\resizebox{!}{4em}{\begin{quantikz}[baseline={([yshift={-7\ht\strutbox}]current bounding box.north)}] \lstick{$\ket{0}$} & \gate{R_y(x)}\gategroup[1,steps=1,style={dashed, rounded corners,inner xsep=2pt}, background,label style={label position=below,anchor=north,yshift=-0.2cm}]{{E}} & \gate{R_x(\theta)} & \gate{H} & \qw \end{quantikz}}}

\sbox4{\resizebox{!}{4em}{\begin{quantikz}[baseline={([yshift={-7\ht\strutbox}]current bounding box.north)}] \lstick{$\ket{0}$} & \gate{R_y(x)}\gategroup[1,steps=1,style={dashed, rounded corners,inner xsep=2pt}, background,label style={label position=below,anchor=north,yshift=-0.2cm}]{{E}} & \gate{R_x(\theta)} & \gate{R_y(x)}\gategroup[1,steps=1,style={dashed, rounded corners,inner xsep=2pt}, background,label style={label position=below,anchor=north,yshift=-0.2cm}]{{E}} & \gate{R_x(\theta)} & \qw \end{quantikz}}}

\sbox5{\resizebox{!}{4em}{\begin{quantikz}[baseline={([yshift={-7\ht\strutbox}]current bounding box.north)}] \lstick{$\ket{0}$} & \gate{R_x(\theta)} & \gate{H} & \gate{R_y(x)}\gategroup[1,steps=1,style={dashed, rounded corners,inner xsep=2pt}, background,label style={label position=below,anchor=north,yshift=-0.2cm}]{{E}} & \gate{H} & \gate{R_x(\theta)} & \qw \end{quantikz}}}

\sbox6{\resizebox{15em}{!}{\begin{quantikz}[baseline={([yshift={-12\ht\strutbox}]current bounding box.north)}] \lstick{$\ket{0}$} & \gate{R_y(x)}\gategroup[1,steps=1,style={dashed, rounded corners,inner xsep=2pt}, background,label style={label position=below,anchor=north,yshift=-0.2cm}]{{E}} & \gate{R_x(\theta)} & \gate{H} & \push{...}\\
\push{...} &\qw & \gate{R_y(x)}\gategroup[1,steps=1,style={dashed, rounded corners,inner xsep=2pt}, background,label style={label position=below,anchor=north,yshift=-0.2cm}]{{E}} & \gate{R_x(\theta)} & \gate{H} & \qw \end{quantikz}}}

\begin{table*}[p]
    \centering
    \begin{tabular}{lccc}
        \textbf{PQC} & \textbf{Fluctiation of expectation values} & \textbf{Amplitudes} & \textbf{Phases} \\
        \usebox0 & 
        \includegraphics[width=0.25\textwidth]{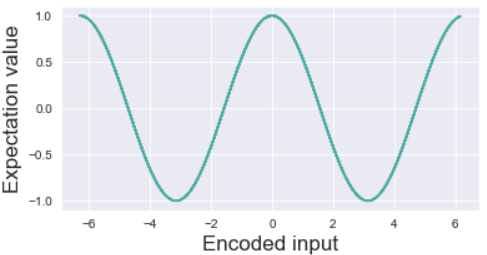} & \includegraphics[width=0.14\textwidth]{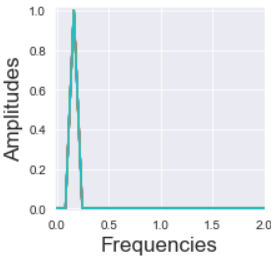} & \includegraphics[width=0.14\textwidth]{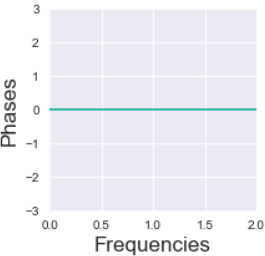}\\
        \usebox1 & 
        \includegraphics[width=0.25\textwidth]{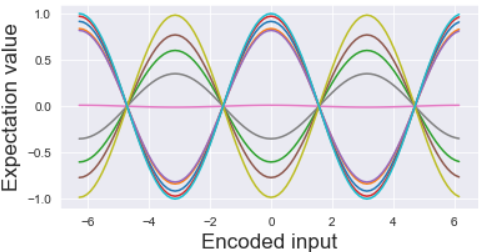} & \includegraphics[width=0.14\textwidth]{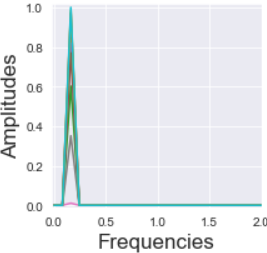} & \includegraphics[width=0.14\textwidth]{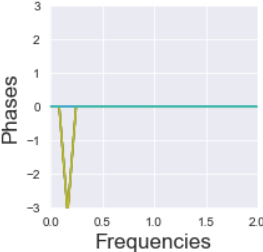}\\
        \usebox2 & 
        \includegraphics[width=0.25\textwidth]{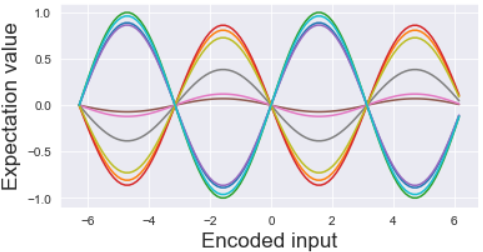} & \includegraphics[width=0.14\textwidth]{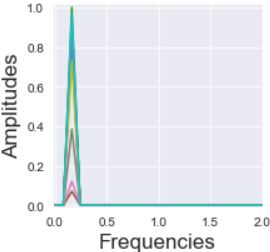} & \includegraphics[width=0.14\textwidth]{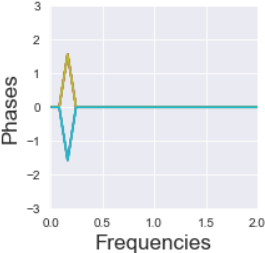}\\
        \usebox3 & 
        \includegraphics[width=0.25\textwidth]{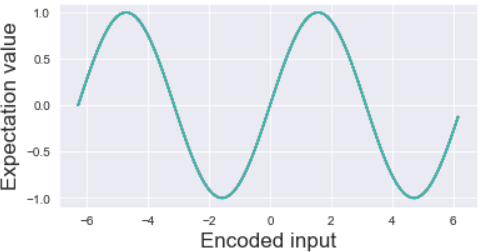} & \includegraphics[width=0.14\textwidth]{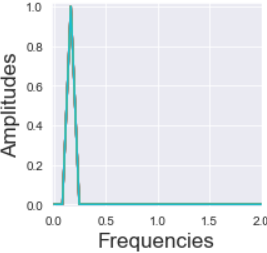} & \includegraphics[width=0.14\textwidth]{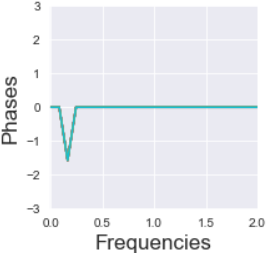}\\
        \usebox4 & 
        \includegraphics[width=0.25\textwidth]{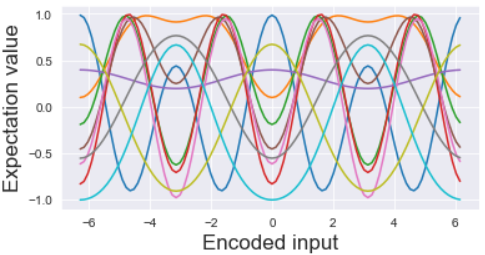} & \includegraphics[width=0.14\textwidth]{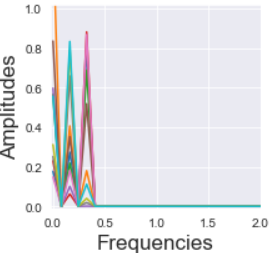} & \includegraphics[width=0.14\textwidth]{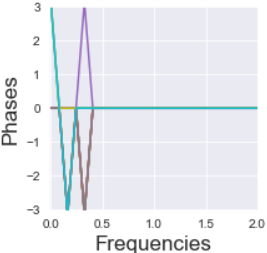}\\
        \usebox5 & 
        \includegraphics[width=0.25\textwidth]{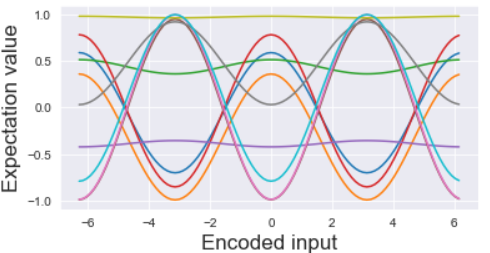} & \includegraphics[width=0.14\textwidth]{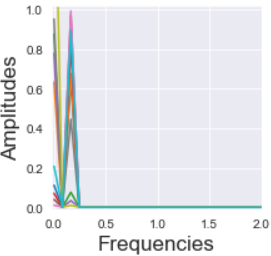} & \includegraphics[width=0.14\textwidth]{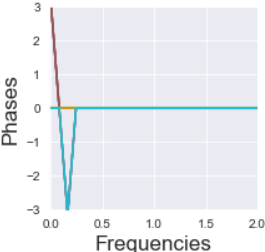}\\
        \usebox6 & 
        \includegraphics[width=0.25\textwidth]{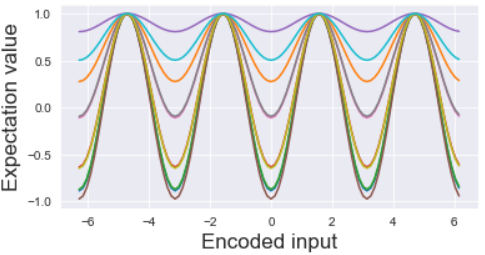} & \includegraphics[width=0.14\textwidth]{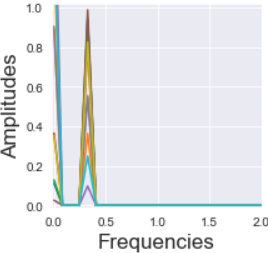} & \includegraphics[width=0.14\textwidth]{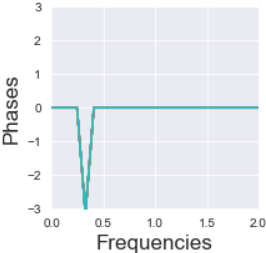}\\
    \end{tabular}
    \caption{\small{Visualization of a dependence between circuit embedding complexity and heterogeneity of functions that it can represent. The left most column represents the exemplary circuit architectures, for which the embedding layers are fixed to be $R_y$ gates (highlighted by dotted boxes). The embedding can either be preceded by preparatory gates or be repeated, which is then followed by post-processing gates. The effect of these modifications are shown in the next column, which shows fluctuations of expectation values for different input values in range $[-2 \pi, 2 \pi]$. $\theta$ for parametrizable gates were sampled at random $\theta \sim U[-\pi, \pi]$ and the experiment was repeated 10 times. It is desirable to see a vast variety of functions plotted in this column as this implies that the circuit's parameters can be tweaked in a way to represent a considerable family of functions. The two columns on the right display amplitude and phase spectrums acquired by Fourier transformation.}}
    \label{tab:fourier_beauty}
\end{table*}
}
     


\end{document}